\documentclass[useAMS,usenatbib]{mn2e} 
\pdfoutput=1
\usepackage{aas_macros}
\usepackage{graphics}
\usepackage{times}
\usepackage{amsmath,amssymb}
\usepackage{bm}

\def\be{\begin{equation}}
\def\ee{\end{equation}}
\def\ba{\begin{eqnarray}}
\def\ea{\end{eqnarray}}

\def\lsim{\mathrel{\rlap{\lower4pt\hbox{\hskip1pt$\sim$}}
    \raise1pt\hbox{$<$}}}                
\def\gsim{\mathrel{\rlap{\lower4pt\hbox{\hskip1pt$\sim$}}
    \raise1pt\hbox{$>$}}}

\pagerange{\pageref{firstpage}--\pageref{lastpage}}
\pubyear{2015}

\begin{document}

\label{firstpage}

\title[Separating weak lensing and intrinsic alignments]{Separating weak lensing and intrinsic
  alignments using radio observations}
\author[Lee Whittaker, Michael L. Brown \& Richard A. Battye]{Lee Whittaker,
  Michael L. Brown, \& Richard A. Battye\\ Jodrell Bank Centre for Astrophysics,
  School of Physics and Astronomy, University of Manchester, Oxford
  Road, Manchester M13 9PL} \date{\today}

\maketitle

\begin{abstract}
We discuss methods for performing weak lensing using radio observations to recover information about the intrinsic structural properties of the source galaxies. Radio surveys provide unique information that can benefit weak lensing studies, such as HI emission, which may be used to construct galaxy velocity maps, and polarized synchrotron radiation; both of which provide information about the unlensed galaxy and can be used to reduce galaxy shape noise and the contribution of intrinsic alignments. Using a proxy for the intrinsic position angle of an observed galaxy, we develop techniques for cleanly separating weak gravitational lensing signals from intrinsic alignment contamination in forthcoming radio surveys. Random errors on the intrinsic orientation estimates introduce biases into the shear and intrinsic alignment estimates. However, we show that these biases can be corrected for if the error distribution is accurately known. We demonstrate our methods using simulations, where we reconstruct the shear and intrinsic alignment auto and cross-power spectra in three overlapping redshift bins. We find that the intrinsic position angle information can be used to successfully reconstruct both the lensing and intrinsic alignment power spectra with negligible residual bias.
\end{abstract}

\begin{keywords}
  methods: statistical - methods: analytical - cosmology: theory - weak gravitational lensing
\end{keywords}

\section{Introduction}
\label{sec:intro}
Gravitational lensing describes how light from a distant background object is deflected due to the curvature of spacetime induced by the presence of a foreground matter and energy distribution. It is a rich phenomenon and can manifest itself in many ways. Two commonly studied regimes are ``strong lensing'' and ``weak lensing''. Strong lensing is concerned with the distortion of a set of background galaxies or galaxy clusters due to a large foreground mass distribution close to the line of sight. This regime results in multiple images of galaxies (first detected with the observation of the Twin Quasar \citep{Walsh.1979}) or, under the right conditions, Einstein rings. Weak lensing is concerned with the small but coherent distortions of a set of background objects and, as such, is difficult to detect \citep{kaiser93}.

The distortion of an image can be described by a linearized mapping that can be decomposed into two components - convergence and shear (see e.g. \cite{Bartelmann.2001}). On cosmological scales one is primarily interested in measuring the convergence power spectrum, which can be reconstructed using estimates of the shear. On these scales the shear is a result of the large-scale structure of the Universe and is known as cosmic shear. In 2000 cosmic shear was detected by four independent groups (\citealt{bacon00, kaiser00, vanwaerbeke00, wittman00}). These detections laid the foundations for observational weak lensing, which has been increasing in precision ever since and has now become a promising tool for probing cosmology \citep{kilbinger13}.

The standard method for performing a cosmic shear measurement requires averaging over the observed ellipticities of a sufficient number of background galaxies and assuming that the average ellipticity is the consequence of cosmic shear. This method is built on the assumption that there is a zero intrinsic alignment (IA) of the source galaxies. However, non-zero IA signals were predicted as early as 2000 (\citealt{heavens00, croft00, crittenden01, catelan01}), with a subsequent detection being made in low-redshift surveys soon after this \citep{brown02}. The presence of IA has the effect of producing a false shear signal and hence a bias in the standard method. Although it has not had a significant impact on the present generation of surveys (e.g. \cite{heymans12}), it is likely to be important in forthcoming surveys such as those performed by, for example, DES \citep{des05}, Euclid \citep{cimatti12} and the LSST \citep{lsst12}. 

Most weak lensing surveys so far have been performed in the optical waveband, but the SKA promises the possibility of performing surveys in the radio waveband \citep{brown15}. Such a survey offers some unique advantages. Firstly, based on an idea originally proposed by \cite{blain02}, \cite{morales06} introduced a method for performing weak lensing using resolved galaxy velocity maps. Radio HI emission is the most promising part of the electromagnetic spectrum to construct the velocity maps, due to the brightness of the emission lines and the well understood luminosity characteristics. Gravitational lensing leads to a velocity map which is inconsistent with the observed galaxy image and \cite{morales06} showed that this effect can be used to recover estimates of the underlying shear signal. In principle, this method removes the contribution of both galaxy shape noise and the effects of intrinsic alignments from weak lensing surveys. It does, however, require velocity maps from well resolved galaxy images, which reduces the number density of available galaxies in the survey. However, using a toy model \cite{morales06} showed that this method may be competitive in future radio surveys, such as with the SKA.

Secondly, \cite{brown11a} (hereafter BB11) suggested a new technique in radio weak lensing which would use the polarization information contained in the radio emission of a source galaxy as a tracer for the intrinsic position angle (IPA) of the galaxy. It had previously been shown \citep{Dyer.1992} that the net polarization position angle is unaffected by a gravitational lens. For future deep radio surveys, the population of observed galaxies is expected to be dominated by star forming galaxies. The dominant source of radio emission from such a galaxy is expected to be synchrotron radiation emitted as electrons are accelerated by the large scale magnetic fields within that galaxy. This emission gives rise to a net polarization position angle (PPA) which, on average, is anti-aligned with the plane of the galaxy \citep{stil09}, providing information about the galaxy's intrinsic orientation. It was shown that such information can be used to construct a shear estimator which greatly reduces the biases resulting from intrinsic alignments compared to the standard method, and also reduces the errors on the shear estimates. It was shown \citep{brown11b} that this new method can be used to create foreground mass reconstructions with accuracies comparable with the standard method, subject to specific assumptions on the size of the errors on the estimates of the intrinsic orientations of the galaxies, and the fraction of galaxies with reliable polarization information. In principle, the method can be applied to estimates of the IPA from any source, and a similar analysis could also be applied to the technique described by \cite{morales06}.

The method displays great promise. However, there is a small residual bias in the estimator when there is both a non-zero error in the IPA estimates and a non zero IA signal. In this paper we develop improved estimators which remove this bias. In Section \ref{sec:method} we present an overview of the method proposed by BB11. We discuss the noise properties of the method and the residual bias which is introduced in the presence of both an IA signal and an error on the intrinsic position angle estimates. We address this bias by introducing a correction term, which depends on the form of the error distribution.

In Section \ref{sec:aa} we extend the angle-only estimator, introduced by \cite{whittaker14}, to include IPA information and we also introduce a hybrid method that combines an angle-only estimate of the intrinsic alignment with full ellipticity information. In Section \ref{sec:sims} we test the methods using simulations by reconstructing the shear and IA auto and cross-power spectra in three overlapping redshift bins. We conclude in Section \ref{sec:conclusion}.

Throughout the paper we assume that we have reliable IPA information for every galaxy for which we have reliable ellipticity or observed position angle measurements. For a real radio survey this would not be the case. However, the purpose of this paper is to demonstrate the potential of the methods presented, provided that we have a sufficient number of galaxies to recover a reliable shear estimate. A detailed discussion of the fraction of galaxies expected to have reliable polarization information can be found in BB11.

\section{Methods and techniques}
\label{sec:method}
Working well within the weak lensing regime we can express the observed ellipticity of a galaxy, $\bm{\epsilon}^{\mathrm{obs}}$, as the sum of the intrinsic ellipticity, $\bm{\epsilon}^{\mathrm{int}}$, the shear, $\bm{\gamma}$, and a measurement error, $\delta\bm{\epsilon}$, such that
\begin{equation}\label{eq:obsellipse}
\boldsymbol{\epsilon}^{\mathrm{obs}}=\bm{\epsilon}^{\mathrm{int}}+\bm{\gamma}+\delta\bm{\epsilon}.
\end{equation}
Considering a small region of the sky, or a cell, such that the shear can be considered constant within that cell, and assuming that the mean intrinsic ellipticity of the galaxies in that particular cell is zero, we can recover an unbiased estimate of the shear by averaging over the observed galaxy ellipticities:
\begin{equation}\label{eq:standardest}
\hat{\bm{\gamma}}=\frac{1}{N}\sum_{i=1}^N\bm{\epsilon}_i^{\mathrm{obs}}.
\end{equation}

If the unlensed galaxies within a particular region of sky are randomly orientated, then the average intrinsic ellipticity is zero. However, there is theoretical motivation (\citealt{catelan01, crittenden01,jing02,mackey02,hirata04}) and observational evidence (\citealt{brown02, heymans04, mandelbaum06, mandelbaum09, hirata07, brainerd09}) indicating that during galaxy formation, correlations between the intrinsic position angles of galaxies, $\alpha^{\mathrm{int}}$, may arise if those galaxies share an evolutionary history. This is the source of the IA signal. If we now assume that the mean intrinsic ellipticity (or, equivalently, that the IA signal) is non-zero, then the standard estimator of equation (\ref{eq:standardest}) is biased, such that
\begin{equation}\label{eq:standardbias}
\left<\hat{\bm{\gamma}}-\bm{\gamma}\right>=\left<\boldsymbol{\epsilon}^{\mathrm{int}}\right>.
\end{equation}
If we assume that the intrinsic ellipticity of a single galaxy can be expressed as the sum of the intrinsic alignment signal, $\bm{\gamma}^{\mathrm{IA}}$, and a randomly orientated ellipticity, $\bm{\epsilon}^{\mathrm{ran}}$, then
\begin{equation}\label{eq:intellip}
\bm{\epsilon}^{\mathrm{int}}=\bm{\gamma}^{\mathrm{IA}}+\bm{\epsilon}^{\mathrm{ran}},
\end{equation}
and the bias in the standard estimator becomes
\begin{equation}\label{eq:standardbias2}
\left<\hat{\bm{\gamma}}-\bm{\gamma}\right>=\bm{\gamma}^{\mathrm{IA}}.
\end{equation}
Hence, an estimate of the shear recovered using the standard method yields a result which is biased by the IA signal.
\subsection{The Brown \& Battye (BB) Estimator}
\label{subsec:B&Best}
In order to mitigate the effects of the bias introduced by IA, BB11 proposed using polarization information from radio surveys to recover an estimate of the IPA, $\hat{\alpha}^{\mathrm{int}}$. It was found that, for the ideal case where there is a zero error on the IPA measurement, and where the PPA is an exact tracer of $\alpha^{\mathrm{int}}$, the shear can be recovered exactly, using only two source galaxies. 

\begin{figure*}
\begin{minipage}{6in}
\includegraphics{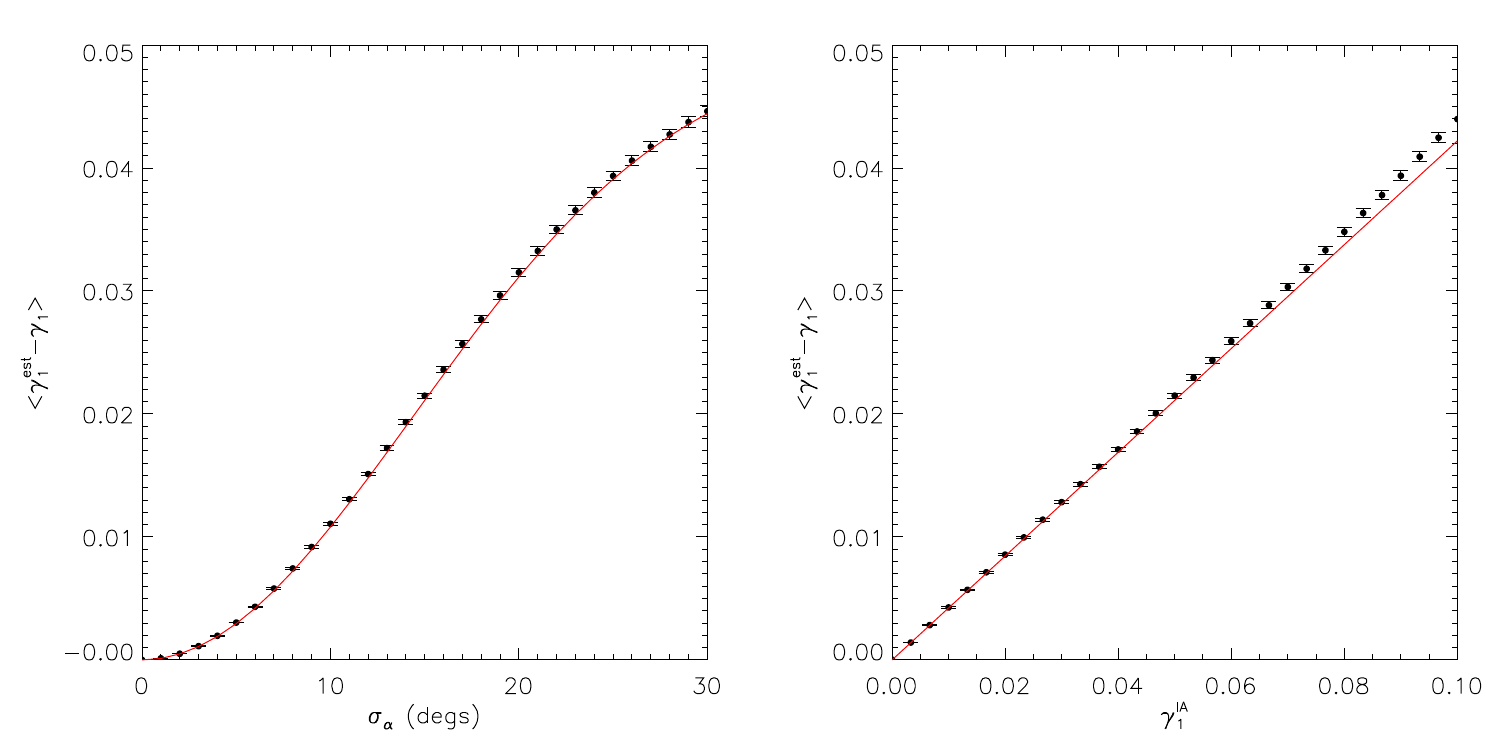}
\caption{The residual bias in the BB estimator from $10^4$ realizations, with $10^4$ galaxies in each realization. For each realization $\gamma_1$, $\gamma_2$ and $\gamma_2^{\mathrm{IA}}$ are selected randomly with a range $[-0.1,0.1]$. The \emph{left panel} shows the bias in the BB estimator as a function of $\sigma_{\alpha^{\mathrm{int}}}$, with $\gamma_1^{\mathrm{IA}}=0.05$. The \emph{right panel} shows the bias as a function of $\gamma_1^{\mathrm{IA}}$ with $\sigma_{\mathrm{int}}=15^{\circ}$. In both cases the red curve is the linear approximation of the bias, given in equation (\ref{eq:lin_bias}).}
\label{fig:bb_bias}
\end{minipage}
\end{figure*}

Expressing the intrinsic ellipticity in polar coordinates, the components of the observed ellipticity can be written as
\begin{align}\label{eq:obsellip_intpol}
\epsilon_1^{\mathrm{obs}}=&\left|\bm{\epsilon}^{\mathrm{int}}\right|\cos\left(2\alpha^{\mathrm{int}}\right)+\gamma_1+\delta\epsilon_1,\nonumber\\
\epsilon_2^{\mathrm{obs}}=&\left|\bm{\epsilon}^{\mathrm{int}}\right|\sin\left(2\alpha^{\mathrm{int}}\right)+\gamma_2+\delta\epsilon_2.
\end{align}
If we define the pseudo-vector
\begin{equation}\label{eq:pseudodef}
\hat{\bm{n}}_i=\left(
\begin{array}{c}
\sin\left(2\hat{\alpha}_i^{\mathrm{int}}\right)\\
-\cos\left(2\hat{\alpha}_i^{\mathrm{int}}\right)
\end{array}\right),
\end{equation}
where $\hat{\alpha}^{\mathrm{int}}$ is an estimate of the intrinsic position angle provided by a measurement of the IPA, a new estimator for the shear can be derived, such that
\begin{equation}\label{eq:Brownest}
\hat{\boldsymbol{\gamma}}=\mathbf{A}^{-1}\boldsymbol{b},
\end{equation}
where $\mathbf{A}$ is a $2\times2$ matrix and $\boldsymbol{b}$ is a two-component vector
\begin{equation}\label{eq:mat_A}
\mathbf{A}=\sum_{i=1}^Nw_i\hat{\boldsymbol{n}}_i\hat{\boldsymbol{n}}_i^T,
\end{equation}
\begin{equation}\label{eq:vec_b}
\boldsymbol{b}=\sum_{i=1}^Nw_i\left(\boldsymbol{\epsilon}^{\mathrm{obs}}_i\cdot\hat{\boldsymbol{n}}_i\right)\hat{\boldsymbol{n}}_i,
\end{equation}
and $w_i$ is a normalized arbitrary weight assigned to each galaxy.
 
In the presence of a non-zero IA signal and a non-zero error on the estimate of $\alpha^{\mathrm{int}}$, it is found that the estimator given in equation (\ref{eq:Brownest}) is biased, although this bias is suppressed significantly with respect to that of the standard estimator. If we assume that the components of the intrinsic ellipticity are isotropically distributed about the IA vector, $\boldsymbol{\gamma}^{\mathrm{IA}}$, and use a uniform weighting, such that $w_i=1$, it is possible to gain some insight into the nature of this bias. If we make the further assumptions that $N\gg1$ and the IA signal is much smaller than the spread in intrinsic ellipticities, $\sigma_{\epsilon}$, that is if $\left|\bm{\gamma}\right|^{\mathrm{IA}}\ll\sigma_{\epsilon}$, we can approximate $\mathbf{A}$ to leading order in $\boldsymbol{\gamma}^{\mathrm{IA}}$ as
\begin{equation}\label{eq:approxA}
\mathbf{A}\approx\frac{N}{2}\mathbf{I},
\end{equation} 
and hence the estimator can then be approximated as
\begin{equation}\label{eq:brownestapprox}
\hat{\boldsymbol{\gamma}}\approx\frac{2}{N}\boldsymbol{b}.
\end{equation}
The noise properties inherent in using measurements of the PPA as a tracer of $\alpha^{\mathrm{int}}$ are discussed in BB11. For this discussion we assume that the measurement error, $\delta\alpha^{\mathrm{int}}$, is independent of the true IPA and distributed symmetrically about zero. If we then make the substitution $\hat{\alpha}^{\mathrm{int}}=\alpha^{\mathrm{int}}+\delta\alpha^{\mathrm{int}}$, we can write the expectation value of the trigonometric functions of the IPA as \citep{whittaker14}
\begin{align}\label{eq:unbiasedPA}
\left<\cos\left(2\hat{\alpha}^{\mathrm{int}}\right)\right>=&\left<\cos\left(2\alpha^{\mathrm{int}}\right)\right>\beta_2^{\mathrm{int}},\nonumber\\
\left<\sin\left(2\hat{\alpha}^{\mathrm{int}}\right)\right>=&\left<\sin\left(2\alpha^{\mathrm{int}}\right)\right>\beta_2^{\mathrm{int}},
\end{align}
where
\begin{equation}\label{eq:gen_beta}
\beta_n^{\mathrm{int}}\equiv\left<\cos\left(n\delta\alpha^{\mathrm{int}}\right)\right>,
\end{equation}
which is the mean cosine of the distribution of $\delta\alpha^{\mathrm{int}}$, and where $n$ is an integer. For a Gaussian measurement error, this can be simplified to
\begin{equation}\label{eq:Gauss_beta_n}
\beta_n^{\mathrm{int}}=\exp\left(-\frac{n^2}{2}\sigma_{\alpha^{\mathrm{int}}}^2\right),
\end{equation}
where $\sigma_{\alpha^{\mathrm{int}}}$ is the standard deviation of the measurement error and is expressed in radians. Taking the limit $N\rightarrow\infty$ and using the result of equation (\ref{eq:unbiasedPA}), it can be shown that equation (\ref{eq:brownestapprox}) may be expanded to first order in the shear and IA, such that the bias in the estimator is
\begin{equation}\label{eq:lin_bias}
\left<\hat{\boldsymbol{\gamma}}-\bm{\gamma}\right>\approx\left(1-\beta_2^{\mathrm{int}}\right)\bm{\gamma}^{\mathrm{IA}}.
\end{equation}
For $\sigma_{\alpha^{\mathrm{int}}}\ll1$, one finds that $\left<\hat{\bm{\gamma}}-\bm{\gamma}\right>\approx2\sigma_{\alpha^{\mathrm{int}}}^2\bm{\gamma}^{\mathrm{IA}}$ and, therefore, we see that the bias is suppressed by a factor of $2\sigma_{\alpha^{\mathrm{int}}}^2$ relative to the standard estimator. The bias in the BB estimator is illustrated in Figure \ref{fig:bb_bias}, where we assume a Gaussian measurement error on the estimate of $\alpha^{\mathrm{int}}$ and a Rayleigh distribution for the intrinsic ellipticity distribution (which we define as the distribution of $\left|\bm{\epsilon}^{\mathrm{ran}}\right|$), such that
\begin{equation}\label{eq:gauss_dist}
f\left(\left|\bm{\epsilon}^{\mathrm{ran}}\right|\right)=\frac{\left|\bm{\epsilon}^{\mathrm{ran}}\right|}{\sigma_{\epsilon}^2\left(1-\exp\left(-\frac{\left|\bm{\epsilon}_{\mathrm{max}}^{\mathrm{ran}}\right|^2}{2\sigma_{\epsilon}^2}\right)\right)}\exp\left(-\frac{\left|\bm{\epsilon}^{\mathrm{ran}}\right|^2}{2\sigma_{\epsilon}^2}\right),
\end{equation}
where $\left|\bm{\epsilon}^{\mathrm{ran}}_{\mathrm{max}}\right|$ is the maximum allowed value of the modulus of $\bm{\epsilon}^{\mathrm{ran}}$; for all of the simulations in this paper we have assumed a Rayleigh distribution for $\left|\bm{\epsilon}^{\mathrm{ran}}\right|$ with values of $\left|\bm{\epsilon}^{\mathrm{ran}}_{\mathrm{max}}\right|=1$ and $\sigma_{\epsilon}=0.3/\sqrt{2}$. From Figure \ref{fig:bb_bias} we see that the estimator successfully reduces the bias introduced by the IA. There is, however, a residual bias introduced when both the measurement error on $\alpha^{\mathrm{int}}$ and the IA signal are non-zero.

In the limit $\left|\gamma^{\mathrm{IA}}\right|\ll\sigma_{\epsilon}$ the standard error for the shear estimator can be written as
\begin{equation}\label{eq:standarderrbrown}
\sigma_{\hat{\gamma}}\approx\left[\frac{2\sigma_{\epsilon}^2\left(1-\beta_4^{\mathrm{int}}\right)+2\sigma^2}{N}\right]^{\frac{1}{2}},
\end{equation}
where $\sigma$ is the measurement error on the components of $\bm{\epsilon}^{\mathrm{obs}}$. Assuming this error to be zero, and assuming $\sigma_{\alpha^{\mathrm{int}}}\ll1$, the error can be approximated as
\begin{equation}\label{eq:err_BB11}
\sigma_{\hat{\gamma}}\approx\frac{4\sigma_{\alpha_{\mathrm{int}}}\sigma_{\epsilon^{\mathrm{int}}}}{\sqrt{N}},
\end{equation}
and hence we see that $\sigma_{\hat{\gamma}}$ is suppressed by a factor of $4\sigma_{\alpha^{\mathrm{int}}}$ relative to the standard estimator, in agreement with the findings of BB11.

Given an estimate of the shear, and assuming that the effects of the intrinsic ellipticity can be modeled using equation (\ref{eq:intellip}), an estimate of the IA signal can be recovered trivially, such that
\begin{equation}\label{eq:bb_IA}
\hat{\bm{\gamma}}^{\mathrm{IA}}=\left(\frac{1}{N}\sum_{i=1}^N\bm{\epsilon}_i^{\mathrm{obs}}\right)-\hat{\bm{\gamma}}.
\end{equation}
The bias in the estimate of the IA signal arises from the bias in the shear estimator and hence to first order, has the same magnitude as the bias given in equation (\ref{eq:lin_bias}), but with the opposite sign.
 
To first order in the shear and intrinsic alignment, the error on the IA estimator is due to the random shape noise,
\begin{equation}\label{eq:disp_IA_bb}
\sigma_{\hat{\bm{\gamma}}^{\mathrm{IA}}}=\frac{\sigma_{\mathrm{\epsilon}}}{\sqrt{N}}.
\end{equation}
From equation (\ref{eq:disp_IA_bb}) we see that the error on the IA estimator is independent of the error on $\alpha^{\mathrm{int}}$ and therefore, there is no suppression of this error by $\sigma_{\alpha^{\mathrm{int}}}$. 

\subsection{The Corrected BB (CBB) estimator}
\label{subsec:cbb}
It is possible to construct an unbiased shear estimator in the limit $N\rightarrow\infty$ by following the approach outlined in BB11. This corrected form of the BB estimator (hereafter the CBB estimator) can be written as 
\begin{equation}\label{eq:correctBB}
\hat{\bm{\gamma}}=\mathbf{D}^{-1}\bm{h},
\end{equation}
where $\mathbf{D}$ is a $2\times2$ matrix
\begin{equation}\label{eq:D_matrix}
\mathbf{D}=\sum_{i=1}^N\mathbf{M}_i,
\end{equation}
and where $\boldsymbol{h}$ is a 2-component vector
\begin{equation}\label{eq:h_matrix}
\bm{h}=\sum_{i=1}^N\mathbf{M}_i\bm{\epsilon}_i^{\mathrm{obs}}.
\end{equation}
In Appendix A it is shown that the matrix $\mathbf{M}_i$ is given by
\begin{equation}\label{eq:matrix_M_corr}
\mathbf{M}_i=\left(
\begin{array}{cc}
\beta_4^{\mathrm{int}} - \cos\left(4\hat{\alpha}_{i}^{\mathrm{int}}\right) & -\sin\left(4\hat{\alpha}_{i}^{\mathrm{int}}\right) \\
-\sin\left(4\hat{\alpha}_{i}^{\mathrm{int}}\right) & \beta_4^{\mathrm{int}} + \cos\left(4\hat{\alpha}_{i}^{\mathrm{int}}\right)
\end{array} \right),
\end{equation}
where the term $\beta_4^{\mathrm{int}}$ is defined in equation (\ref{eq:gen_beta}) and corrects for the bias on the trigonometric functions introduced by a measurement error on $\alpha^{\mathrm{int}}$. Once one has an estimate of $\bm{\gamma}$, an estimate of the IA can be recovered using equation (\ref{eq:bb_IA}).

\begin{figure*}
\begin{minipage}{6in}
\includegraphics{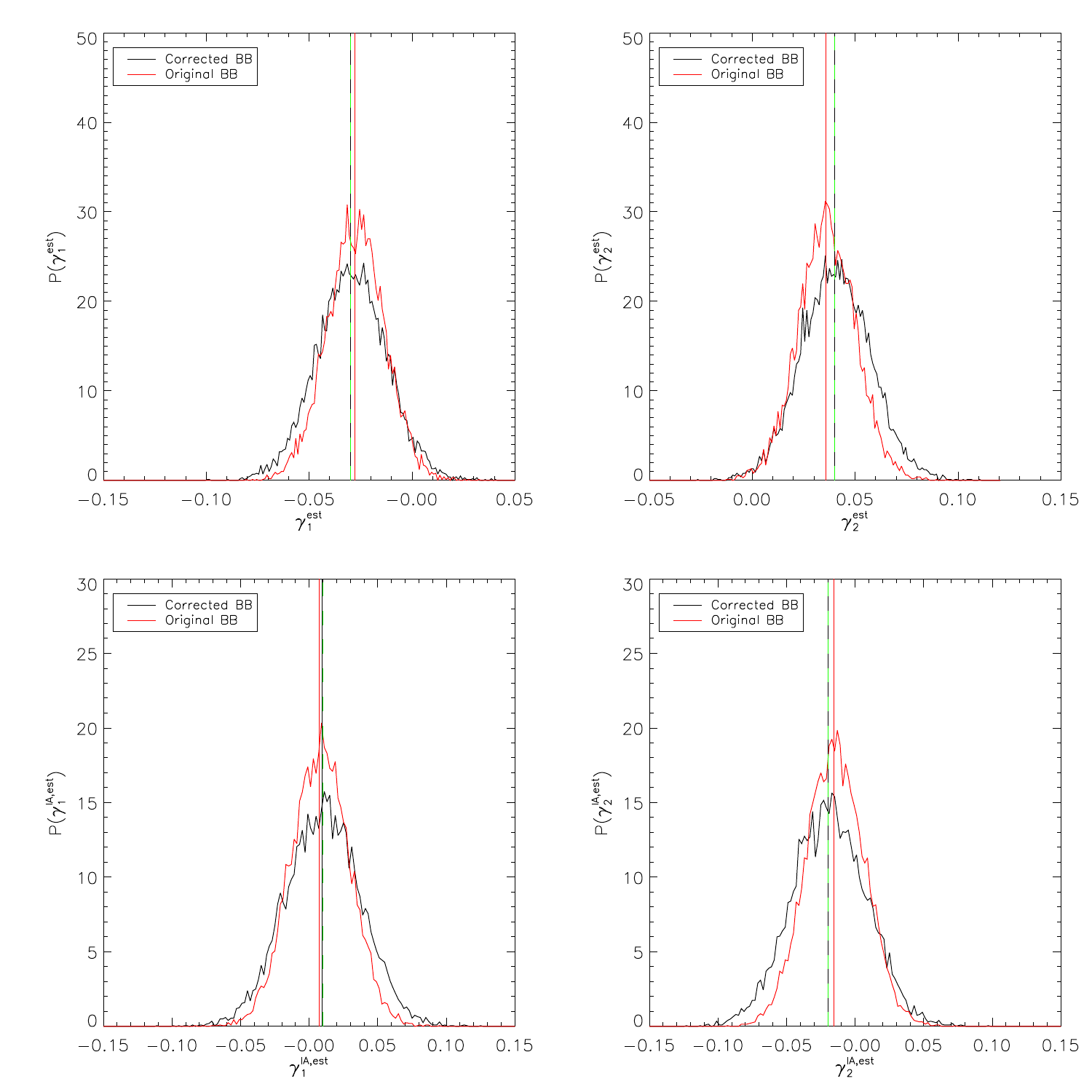}
\caption{The recovered shear and IA estimates from $10^4$ realizations, with each realization consisting of 100 galaxies, and assuming a measurement error on the IPA of $\sigma_{\alpha^{\mathrm{int}}}=10^{\circ}$. The black curves show the distributions of recovered shear and IA estimates when using the CBB estimator (equations (\ref{eq:bb_IA}) and (\ref{eq:correctBB})) and the vertical black lines show the mean recovered shear estimates when using this estimator. The red curves show the distributions of recovered shear estimates when using the original BB estimator (equations (\ref{eq:Brownest}) and (\ref{eq:bb_IA})) and the vertical red line shows the mean recovered shear estimates when using this estimator. The green dashed lines, which lie on top of the black lines, show the input shear signal. The success of the correction to the original BB estimator is clearly visible in these plots. There is, however, a modest increase ($\sim$20\%) in the dispersion of the shear estimates and a $\sim$30\% increase in the dispersion of the IA estimates when using the corrected form of the estimator with this set of input values; this is quantified in Table \ref{table:comp_bb_cbb}.}
\label{fig:disp_bb_cbb}
\end{minipage}
\end{figure*}
\begin{table*}
\begin{minipage}{6in}
\centering
\begin{tabular}{|c|c|c|c|c|c|c|c|c|}
\hline
Estimator $\left(\times10^{-2}\right)$ & $\sigma_{\hat{\gamma}_1}$ & $\sigma_{\hat{\gamma}_2}$ & $\left<\hat{\gamma}_1\right>$ & $\left<\hat{\gamma}_2\right>$ & $\sigma_{\hat{\gamma}_1^{\mathrm{IA}}}$ & $\sigma_{\hat{\gamma}_2^{\mathrm{IA}}}$ & $\left<\hat{\gamma}_1^{\mathrm{IA}}\right>$ & $\left<\hat{\gamma}_2^{\mathrm{IA}}\right>$ \\[0.5ex]
\hline
Original BB & $1.40$ & $1.40$ & $-2.79\pm0.01$ & $3.58\pm0.01$ & $2.13$ & $2.12$ & $0.71\pm0.02$ & $-1.55\pm0.02$ \\
Corrected BB & $1.71$ & $1.70$ & $-3.00\pm0.02$ & $4.02\pm0.02$ & $2.75$ & $2.73$ & $0.92\pm0.03$ & $-1.99\pm0.03$ \\ [1ex]
\hline
\end{tabular}
\caption{The mean and standard deviation of the shear and IA estimates
  recovered from $10^4$ simulations. Values are quoted for both the 
  original BB estimator (equations (\ref{eq:Brownest}) and (\ref{eq:bb_IA})) and the 
  CBB estimator (equations (\ref{eq:bb_IA})) and (\ref{eq:correctBB}). The input shear and IA
  values are $\gamma_1=-0.03$, $\gamma_2=0.04$, $\gamma_1^{\mathrm{IA}}=0.01$ and $\gamma_2^{\mathrm{IA}}=-0.02$.}
\label{table:comp_bb_cbb}
\end{minipage}
\end{table*}
We have tested the performance of the CBB estimator using simulations composed of 100 galaxies and assuming an input shear signal of $\gamma_1=-0.03$ and $\gamma_2=0.04$, and an input IA signal of $\gamma_1^{\mathrm{IA}}=0.01$ and $\gamma_2^{\mathrm{IA}}=-0.02$. We recovered shear and IA estimates from $10^{4}$ realizations using the original form of the BB estimator (equations (\ref{eq:Brownest}) and (\ref{eq:bb_IA})) and the CBB estimator (equations (\ref{eq:bb_IA}) and (\ref{eq:correctBB})). We assumed a zero error on measurements of $\bm{\epsilon}^{\mathrm{obs}}$ and a Gaussian measurement error with r.m.s. $10^{\circ}$ on $\alpha^{\mathrm{int}}$. The results of this test are shown in Figure \ref{fig:disp_bb_cbb}. Table \ref{table:comp_bb_cbb} presents the mean recovered shear and IA estimates and the standard deviation of the estimates. 

\begin{figure*}
\begin{minipage}{6in}
\includegraphics{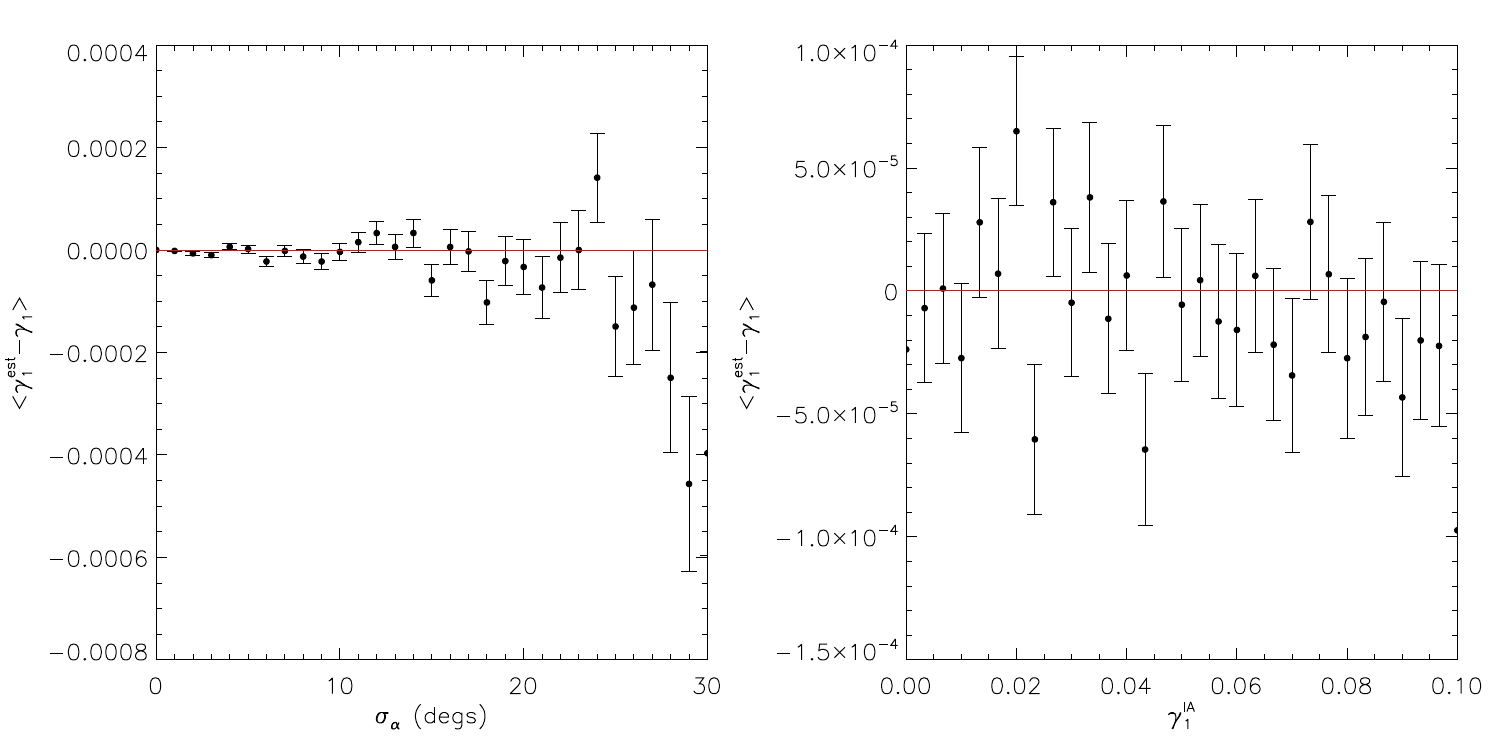}
\caption{Same as for Figure \ref{fig:bb_bias} but for the CBB shear estimator. We see a residual bias which is a result of the finite number of source galaxies, however, this residual bias is much smaller than the residual bias in the original BB estimator, shown in Figure \ref{fig:bb_bias}.}
\label{fig:cbb_bias}
\end{minipage}
\end{figure*}
Figure \ref{fig:cbb_bias} shows the residual bias in the CBB estimator. The bias correction term, $\beta_4^{\mathrm{int}}$, corrects for the bias introduced to the mean trigonometric functions in the BB estimator when there is an error on the estimates of $\alpha^{\mathrm{int}}$. However, for a finite number of source galaxies, there will also be noise in the estimates of the mean trigonometric functions (which enters into the CBB estimator via the inverse of matrix $\mathbf{D}$). This noise propagates nonlinearly into estimates of the shear, resulting in a residual bias which is not corrected for. In the tests we have conducted, we find that this residual bias is always much smaller than the dispersion in the shear estimates and contributes to a negligible residual bias in the power spectra reconstructions discussed in Section \ref{sec:sims}.

In order to estimate the dispersion in the shear and IA estimates, we can write an approximate form of the CBB estimator. To leading order in $\bm{\gamma}$ and $\bm{\gamma}^{\mathrm{IA}}$, the CBB estimator can be written as
\begin{align}\label{eq:lin_cor_bb}
\hat{\gamma}_1=&\frac{1}{N\beta_4^{\mathrm{int}}}\sum_{i=1}^N\biggl[\beta_4^{\mathrm{int}}\epsilon_1^{\mathrm{obs},(i)}-\epsilon_1^{\mathrm{obs},(i)}\cos\left(4\hat{\alpha}_i^{\mathrm{int}}\right)\nonumber\\
&-\epsilon_2^{\mathrm{obs},(i)}\sin\left(4\hat{\alpha}_i^{\mathrm{int}}\right)\biggr], \nonumber \\
\hat{\gamma}_2=&\frac{1}{N\beta_4^{\mathrm{int}}}\sum_{i=1}^N\biggl[\beta_4^{\mathrm{int}}\epsilon_2^{\mathrm{obs},(i)}-\epsilon_1^{\mathrm{obs},(i)}\sin\left(4\hat{\alpha}_i^{\mathrm{int}}\right)\nonumber\\
&+\epsilon_2^{\mathrm{obs},(i)}\cos\left(4\hat{\alpha}_i^{\mathrm{int}}\right)\biggr], \nonumber \\
\hat{\gamma}_1^{\mathrm{IA}}=&\frac{1}{N\beta_4^{\mathrm{int}}}\sum_{i=1}^N\left[\epsilon_1^{\mathrm{obs},(i)}\cos\left(4\hat{\alpha}_i^{\mathrm{int}}\right)+\epsilon_2^{\mathrm{obs},(i)}\sin\left(4\hat{\alpha}_i^{\mathrm{int}}\right)\right], \nonumber \\
\hat{\gamma}_2^{\mathrm{IA}}=&\frac{1}{N\beta_4^{\mathrm{int}}}\sum_{i=1}^N\left[\epsilon_1^{\mathrm{obs},(i)}\sin\left(4\hat{\alpha}_i^{\mathrm{int}}\right)-\epsilon_2^{\mathrm{obs},(i)}\cos\left(4\hat{\alpha}_i^{\mathrm{int}}\right)\right].
\end{align}
The error on the CBB estimator can then be approximated as
\begin{align}\label{eq:corr_err_bb}
\sigma_{\hat{\gamma}_1}=&\sigma_{\hat{\gamma}_2}=\left[\frac{\sigma_{\epsilon}^2\left(1-\beta_4^{\mathrm{int}^2}\right)+\sigma^2\left(1+\beta_4^{\mathrm{int}^2}\right)}{N\beta_4^{\mathrm{int}^2}}\right]^{\frac{1}{2}}, \nonumber \\
\sigma_{\hat{\gamma}_1^{\mathrm{IA}}}=&\sigma_{\hat{\gamma}_2^{\mathrm{IA}}}=\left[\frac{\sigma_{\epsilon}^2+\sigma^2}{N\beta_4^{\mathrm{int}^2}}\right]^{\frac{1}{2}}.
\end{align}

Using equation (\ref{eq:corr_err_bb}) with the input values used to produce Figure \ref{fig:disp_bb_cbb}, we recovered approximate values for the dispersion in the estimators of $\sigma_{\hat{\gamma}_1}=\sigma_{\hat{\gamma}_2}=1.68\times10^{-2}$ and $\sigma_{\hat{\gamma}_1^{\mathrm{IA}}}=\sigma_{\hat{\gamma}_2^{\mathrm{IA}}}=2.71\times10^{-2}$, which are in good agreement with the measured values quoted in Table \ref{table:comp_bb_cbb}. For completeness, we also used equations (\ref{eq:standarderrbrown}) and (\ref{eq:disp_IA_bb}) to recover approximate values for the dispersion in the original BB estimator. These values were $\sigma_{\hat{\gamma}_1}=\sigma_{\hat{\gamma}_2}=1.40\times10^{-2}$ and $\sigma_{\hat{\gamma}_1^{\mathrm{IA}}}=\sigma_{\hat{\gamma}_2^{\mathrm{IA}}}=2.12\times10^{-2}$, which are also in agreement with the values in quoted Table \ref{table:comp_bb_cbb}.

\subsection{Required Galaxy Numbers for the CBB Estimator}
\label{subsec:req_num}
The CBB estimator becomes unstable when there is a low number of background galaxies available in a particular cell. To gain some insight into the source of this issue, we can examine the behaviour of the determinant of matrix $\mathbf{D}$ when a low number of galaxies is considered. The determinant of matrix $\mathbf{D}$ is
\begin{align}\label{eq:det_D}
\mathrm{det}\left(\mathbf{D}\right)=&\beta_4^{\mathrm{int}^2}-\left[\frac{1}{N}\sum_{i=1}^N\cos\left(4\hat{\alpha_i}^{\mathrm{int}}\right)\right]^2\nonumber\\
&-\left[\frac{1}{N}\sum_{i=1}^N\sin\left(4\hat{\alpha_i}^{\mathrm{int}}\right)\right]^2.
\end{align}
As the measurement error on $\alpha^{\mathrm{int}}$ is increased, the bias correction, $\beta_4^{\mathrm{int}}$, decreases. For a finite number of background galaxies, chance alignments of the random components of the intrinsic galaxy orientations can force this determinant to approach zero, with the effect being more likely when the number of galaxies in a cell is low. This, in turn, can produce substantial outliers in the estimated shear values, as the modulus of a particular shear estimate is scaled with the reciprocal of the determinant. It is possible to place constraints on the number of background galaxies required for a reliable shear estimate by assuming that there are enough galaxies in the sample, such that the central limit theorem can be applied to the distributions of the means of the trigonometric functions in equation (\ref{eq:det_D}). We can use this assumption to examine the probability that the sum of the square of the mean trigonometric functions in equation (\ref{eq:det_D}) will lie within a given range of $\beta_4^{\mathrm{int}^2}$. The determinant is independent of the shear and is only dependent on the IA signal at $4^{\text{th}}$ order, so we can safely assume the IA signal to be zero. With these assumptions in place, we choose to constrain the number of galaxies, such that we can exclude values of the reciprocal $>2/\beta_4^{\mathrm{int}^2}$ at a confidence level of $99.99994$\%, which is equivalent to a confidence level of $5\sigma$ for the Gaussian distribution. The choice of $5\sigma$ is selected to mitigate the issue of outliers when considering the simulations in Section \ref{sec:sims}, where we reconstruct the shear and IA auto and cross-power spectra using $\sim$$10^6$ cells per redshift bin and therefore expect typically one cell per reconstruction to have a reciprocal value $>2/\beta_4^{\mathrm{int}^2}$. The constraint on the values of the reciprocal $>2/\beta_4^{\mathrm{int}^2}$ is somewhat arbitrary but serves to provide an upper limit on the dispersion of the shear estimates. 

\begin{figure}
\includegraphics{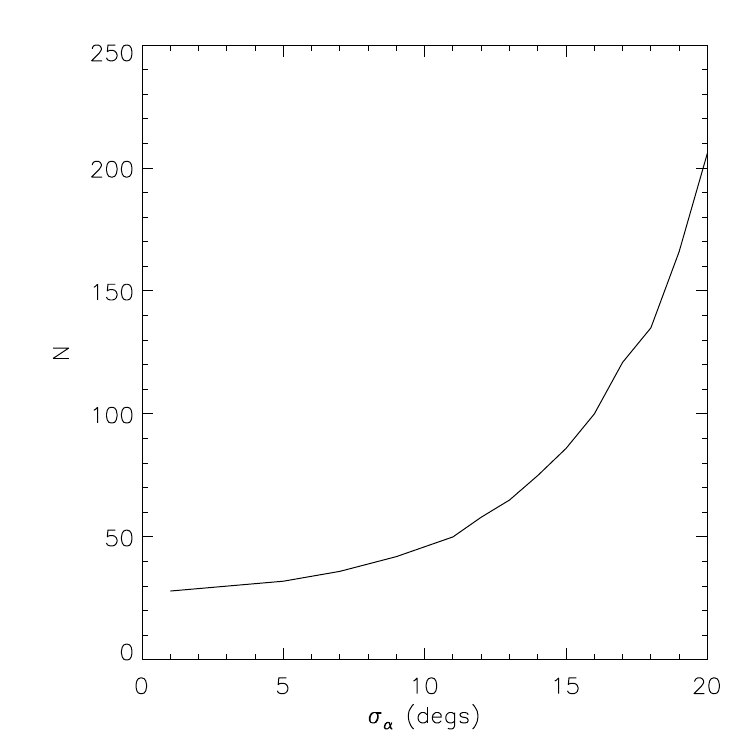}
\caption{The number of galaxies in the sample, as a function of the error on $\alpha^{\mathrm{int}}$, such that the reciprocal of the determinant $<2/\beta_4^{\mathrm{int}^2}$ with a confidence level of $5\sigma$.}
\label{fig:det_betsq}
\end{figure}
This choice of constraint parameters results in Figure \ref{fig:det_betsq}, where we plot the number of galaxies required in the sample as a function of $\sigma_{\alpha^{\mathrm{int}}}$. As an example, let us assume a measurement error on $\alpha^{\mathrm{int}}$ of $10^{\circ}$. Then, from Figure \ref{fig:det_betsq}, we find that we need $\sim$46 galaxies in each cell, such that values of the reciprocal of the determinant $>2/\beta_4^{\mathrm{int}^2}$ are ruled out at a confidence level of $5\sigma$. For Figure \ref{fig:disp_bb_cbb} we considered 100 galaxies per realization and hence outliers were not an issue for these tests. The number density of background galaxies will be fixed for any specific set of observations. However, for a fixed number density of galaxies, the size of the cells may be chosen so that the number of source galaxies within each cell is greater or equal to the number of galaxies required to recover a reliable shear estimate. For a low number density of background galaxies, this will of course result in a large cell size and hence the loss of small scale information.

\section{Alternative Approaches}
\label{sec:aa}
\subsection{Full Angle-Only Estimator (FAO)}
\label{subsec:ao}
In this section we extend the angle-only shear estimator, introduced by \cite{whittaker14}, to include measurements of the IPA. Assuming a prior knowledge of the intrinsic ellipticity distribution, \cite{whittaker14} showed that it is possible to recover an estimate of the shear using only measurements of galaxy position angles. Using measurements of the IPA, as opposed to the observed position angles, this method can be used to recover a direct estimate of the IA signal.  

We begin by writing the IA in polar form, such that
\begin{align}\label{eq:IA_polar}
\gamma_1^{\mathrm{IA}}=&\left|\bm{\gamma}^{\mathrm{IA}}\right|\cos\left(2\alpha^{\mathrm{IA}}\right),\nonumber\\
\gamma_2^{\mathrm{IA}}=&\left|\bm{\gamma}^{\mathrm{IA}}\right|\sin\left(2\alpha^{\mathrm{IA}}\right).
\end{align}
It can then be shown that the means of the cosines and sines of the intrinsic position angles can be written as
\begin{align}\label{eq:mean_cos_sin_PPA}
\left<\cos\left(2\alpha^{\mathrm{int}}\right)\right>=&F_1\left(\left|\bm{\gamma}^{\mathrm{IA}}\right|\right)\cos\left(2\alpha^{\mathrm{IA}}\right),\nonumber\\
\left<\sin\left(2\alpha^{\mathrm{int}}\right)\right>=&F_1\left(\left|\bm{\gamma}^{\mathrm{IA}}\right|\right)\sin\left(2\alpha^{\mathrm{IA}}\right),
\end{align}
where the form of the function $F_1\left(\left|\bm{\gamma}^{\mathrm{IA}}\right|\right)$ is dependent on $f\left(\left|\bm{\epsilon}^{\mathrm{ran}}\right|\right)$ and the assumed model which describes how the IA transforms $\bm{\epsilon}^{\mathrm{ran}}\rightarrow\bm{\epsilon}^{\mathrm{int}}$. Note that $F_1\left(\left|\bm{\gamma}^{\mathrm{IA}}\right|\right)$ is a function of the modulus of the IA only. Assuming a zero measurement error on the intrinsic position angle estimate (i.e. $\hat{\alpha}^{\mathrm{int}}=\alpha^{\mathrm{int}}$) and a sample of $N$ galaxies, we can recover an estimate of the orientation of the IA, such that
\begin{equation}\label{eq:est_aIA}
\hat{\alpha}^{\mathrm{IA}}=\frac{1}{2}\tan^{-1}\left(\frac{\sum_{i=1}^N\sin\left(2\hat{\alpha}_i^{\mathrm{int}}\right)}{\sum_{i=1}^N\cos\left(2\hat{\alpha}_i^{\mathrm{int}}\right)}\right).
\end{equation}
We can also recover an estimate of $F_1\left(\left|\bm{\gamma}^{\mathrm{IA}}\right|\right)$,
\begin{equation}\label{eq:est_F}
F_1\left(\left|\hat{\bm{\gamma}}^{\mathrm{IA}}\right|\right)=\frac{1}{N}\sqrt{\left(\sum_{i=1}^N\cos\left(2\hat{\alpha}_i^{\mathrm{int}}\right)\right)^2+\left(\sum_{i=1}^N\sin\left(2\hat{\alpha}_i^{\mathrm{int}}\right)\right)^2},
\end{equation}
which can be inverted to provide an estimate of $\left|\bm{\gamma}^{\mathrm{IA}}\right|$. Using equations (\ref{eq:est_aIA}) and (\ref{eq:est_F}) we can, therefore, recover a complete estimate of $\bm{\gamma}^{\mathrm{IA}}$ using only measurements of the IPA.

The $F_1\left(\left|\bm{\gamma}^{\mathrm{IA}}\right|\right)$ function describes the mean cosine of the angle between the vectors $\bm{\epsilon}^{\mathrm{ran}}$ and $\bm{\gamma}^{\mathrm{IA}}$ as a function of $\left|\bm{\gamma}^{\mathrm{IA}}\right|$. If we assume that the effects of IA can be modeled using equation (\ref{eq:intellip}), then the $F_1\left(\left|\bm{\gamma}^{\mathrm{IA}}\right|\right)$ function is found to be
\begin{align}\label{eq:F_IA}
F_1\left(\left|\bm{\gamma}^{\mathrm{IA}}\right|\right)=&\frac{1}{\pi}\int_0^{\left|\bm{\epsilon}_{\mathrm{max}}^{\mathrm{ran}}\right|}\int_{-\frac{\pi}{2}}^{\frac{\pi}{2}}\mathrm{d}\alpha^{\mathrm{ran}}\mathrm{d}\left|\bm{\epsilon}^{\mathrm{ran}}\right|f\left(\left|\bm{\epsilon}^{\mathrm{ran}}\right|\right)\nonumber\\
&\times g_1\left(\left|\bm{\gamma}^{\mathrm{IA}}\right|,\left|\bm{\epsilon}^{\mathrm{ran}}\right|,\alpha^{\mathrm{ran}}\right),
\end{align}
where the function $g_1\left(\left|\bm{\gamma}^{\mathrm{IA}}\right|,\left|\bm{\epsilon}^{\mathrm{ran}}\right|,\alpha^{\mathrm{ran}}\right)$ is given as
\begin{equation}\label{eq:g1_F}
g_1\left(\left|\bm{\gamma}^{\mathrm{IA}}\right|,\left|\bm{\epsilon}^{\mathrm{ran}}\right|,\alpha^{\mathrm{ran}}\right)=\frac{\epsilon_1'^2}{\sqrt{\epsilon_1'^2+\epsilon_2'^2}},
\end{equation}
with
\begin{align}\label{eq:edash}
\epsilon_1'=&\left|\bm{\gamma}^{\mathrm{IA}}\right|+\left|\bm{\epsilon}^{\mathrm{ran}}\right|\cos\left(2\alpha^{\mathrm{ran}}\right), \nonumber \\
\epsilon_2'=&\left|\bm{\epsilon}^{\mathrm{ran}}\right|\sin\left(2\alpha^{\mathrm{ran}}\right). \nonumber \\
\end{align}

If we now allow for a measurement error on $\alpha^{\mathrm{int}}$, such that
\begin{equation}\label{eq:errorPPA}
\hat{\alpha}^{\mathrm{int}}=\alpha^{\mathrm{int}}+\delta\alpha^{\mathrm{int}},
\end{equation}
where $\delta\alpha^{\mathrm{int}}$ is independent of the true $\alpha^{\mathrm{int}}$, \cite{whittaker14} show that estimates of $\alpha^{\mathrm{IA}}$ remain unbiased. However, estimates of $\left|\bm{\gamma}^{\mathrm{IA}}\right|$, obtained by inverting the $F_1\left(\bm{\gamma}^{\mathrm{IA}}\right)$ function, become biased. The bias can be corrected for by dividing the $F_1\left(\left|\bm{\gamma}^{\mathrm{IA}}\right|\right)$ function by the correction term $\beta_2^{\mathrm{int}}$, which follows the definition given in equation (\ref{eq:gen_beta}):
\begin{equation}\label{eq:corrected_F}
F_1\left(\left|\hat{\bm{\gamma}}^{\mathrm{IA}}\right|\right)=\frac{1}{N\beta_2^{\mathrm{int}}}\sqrt{\left(\sum_{i=1}^N\cos\left(2\hat{\alpha}_i^{\mathrm{int}}\right)\right)^2+\left(\sum_{i=1}^N\sin\left(2\hat{\alpha}_i^{\mathrm{int}}\right)\right)^2}.
\end{equation}

Assuming that we are working well within the weak lensing regime, such that $\bm{\epsilon}^{\mathrm{obs}}$ can be described using equation (\ref{eq:obsellipse}), and ignoring measurement errors, we can express the observed ellipticity in terms of the shear and IA, such that
\begin{equation}\label{eq:obsellipse_IA}
\bm{\epsilon}^{\mathrm{obs}}=\bm{\gamma}+\bm{\gamma}^{\mathrm{IA}}+\bm{\epsilon}^{\mathrm{ran}}.
\end{equation}
Expressing the observed ellipticity in polar coordinates:
\begin{align}
\epsilon^{\mathrm{obs}}_1=&\left|\bm{\epsilon}^{\mathrm{obs}}\right|\cos\left(2\alpha^{\mathrm{obs}}\right),\nonumber\\
\epsilon^{\mathrm{obs}}_2=&\left|\bm{\epsilon}^{\mathrm{obs}}\right|\sin\left(2\alpha^{\mathrm{obs}}\right),
\end{align}
we can follow the approach described above to recover estimates of the vector $\bm{\gamma}+\bm{\gamma}^{\mathrm{IA}}$ from the observed galaxy orientations. If we assume a measurement error on $\alpha^{\mathrm{obs}}$ which is independent of the true value
\begin{equation}\label{eq:a_obs_err}
\hat{\alpha}^{\mathrm{obs}}=\alpha^{\mathrm{obs}}+\delta\alpha^{\mathrm{obs}},
\end{equation}
we can define the terms $\beta_n^{\mathrm{obs}}$, such that
\begin{equation}\label{eq:beta_obs}
\beta_n^{\mathrm{obs}}\equiv\left<\cos\left(n\delta\alpha^{\mathrm{obs}}\right)\right>.
\end{equation}
Upon expressing the vector $\bm{\gamma}+\bm{\gamma}^{\mathrm{IA}}$ as
\begin{align}
\gamma_1+\gamma^{\mathrm{IA}}_1=&\left|\bm{\gamma}^{\mathrm{tot}}\right|\cos\left(2\alpha^{\mathrm{tot}}\right),\nonumber\\
\gamma_2+\gamma^{\mathrm{IA}}_2=&\left|\bm{\gamma}^{\mathrm{tot}}\right|\sin\left(2\alpha^{\mathrm{tot}}\right),
\end{align}
we can recover an estimate of $\alpha^{\mathrm{tot}}$, such that
\begin{equation}\label{eq:est_atot}
\hat{\alpha}^{\mathrm{tot}}=\frac{1}{2}\tan^{-1}\left(\frac{\sum_{i=1}^N\sin\left(2\hat{\alpha}_i^{\mathrm{obs}}\right)}{\sum_{i=1}^N\cos\left(2\hat{\alpha}_i^{\mathrm{obs}}\right)}\right),
\end{equation}
and an estimate of $\left|\bm{\gamma}^{\mathrm{tot}}\right|$ which satisfies the equation
\begin{equation}\label{eq:corrected_F_tot}
F_1\left(\left|\hat{\bm{\gamma}}^{\mathrm{tot}}\right|\right)=\frac{1}{N\beta_2^{\mathrm{obs}}}\sqrt{\left(\sum_{i=1}^N\cos\left(2\hat{\alpha}_i^{\mathrm{obs}}\right)\right)^2+\left(\sum_{i=1}^N\sin\left(2\hat{\alpha}_i^{\mathrm{obs}}\right)\right)^2},
\end{equation}
and which provides us with an estimate of the vector $\bm{\gamma}^{\mathrm{tot}}$ using measurements of the observed position angle, $\hat{\alpha}^{\mathrm{obs}}$, only.
 
The vector $\bm{\epsilon}^{\mathrm{ran}}$, given in equation (\ref{eq:obsellipse_IA}), is identical to that given in equation (\ref{eq:intellip}), therefore the form of the $F_1\left(\left|\bm{\gamma}^{\mathrm{tot}}\right|\right)$ function, given in equation (\ref{eq:corrected_F_tot}), is identical to the form of the $F_1\left(\left|\bm{\gamma}^{\mathrm{IA}}\right|\right)$ function in equation (\ref{eq:F_IA}), with the substitution $\left|\bm{\gamma}^{\mathrm{IA}}\right|\rightarrow\left|\bm{\gamma}^{\mathrm{tot}}\right|$. The term $\beta_2^{\mathrm{obs}}$ corrects for the bias introduced by the measurement error on $\alpha^{\mathrm{obs}}$. Once we have recovered estimates of $\bm{\gamma}^{\mathrm{IA}}$ and $\bm{\gamma}^{\mathrm{tot}}$, an estimate of the shear can be recovered trivially, such that
\begin{equation}\label{eq:ao_shear}
\hat{\bm{\gamma}}=\hat{\bm{\gamma}}^{\mathrm{tot}}-\hat{\bm{\gamma}}^{\mathrm{IA}}.
\end{equation}

To summarize, the full angle-only estimator (hereafter the FAO estimator) first requires an estimate of the intrinsic ellipticity distribution, $f\left(\left|\bm{\epsilon}^{\mathrm{ran}}\right|\right)$. We can use this information with measurements of the IPA only to recover an estimate of $\bm{\gamma}^{\mathrm{IA}}$ via equations (\ref{eq:est_aIA}) and (\ref{eq:corrected_F}). An estimate of the vector $\bm{\gamma}+\bm{\gamma}^{\mathrm{IA}}$ can also be obtained using the same method via equations (\ref{eq:est_atot}) and (\ref{eq:corrected_F_tot}). Finally, we use equation (\ref{eq:ao_shear}) to recover an estimate of $\bm{\gamma}$.

\begin{figure*}
\begin{minipage}{6in}
\includegraphics{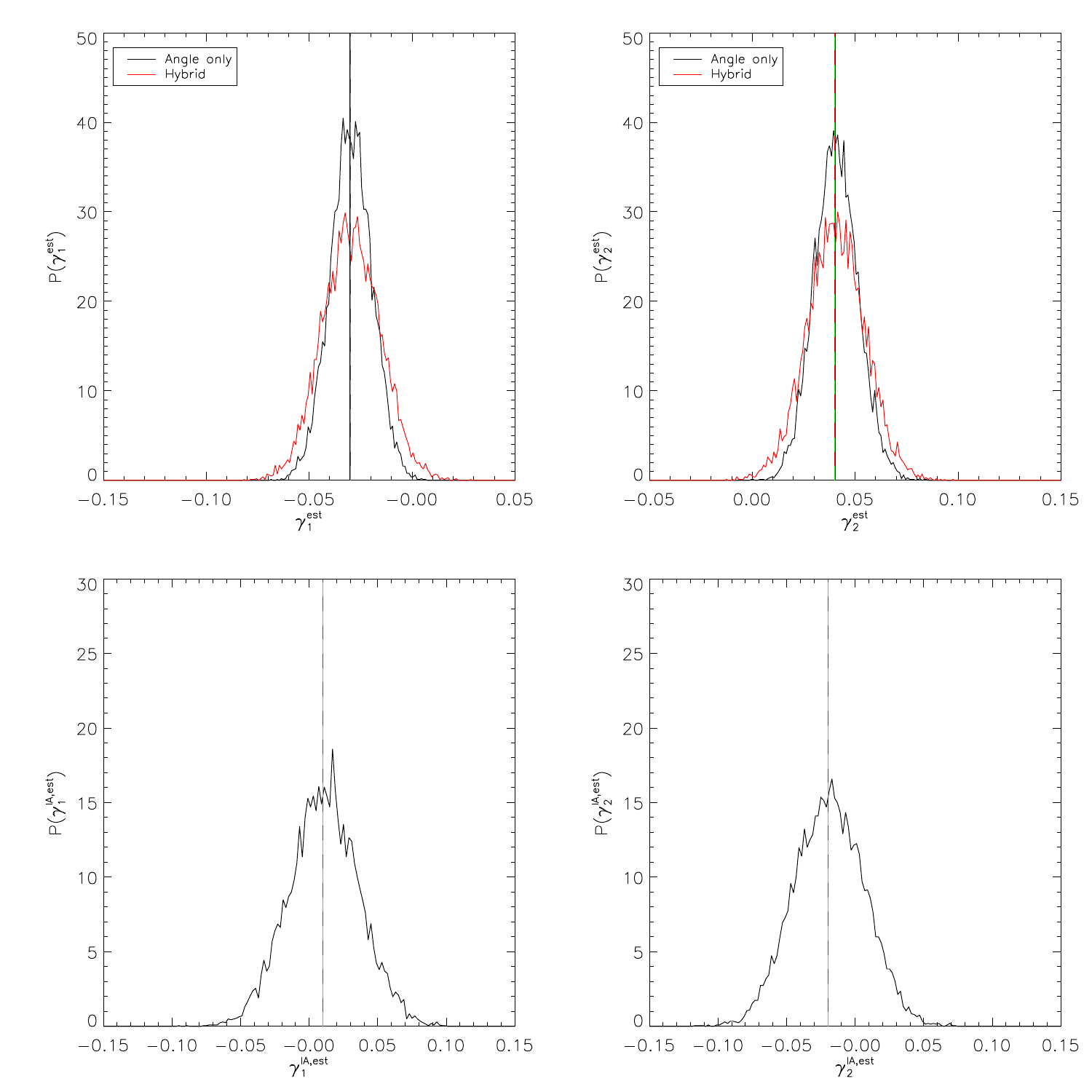}
\caption{The recovered shear and IA estimates from $10^4$ realizations, with each realizations consisting of 100 galaxies, and assuming a measurement error on $\alpha^{\mathrm{int}}$ of $\sigma_{\alpha^{\mathrm{int}}}=10^{\circ}$. The black curves show the distributions of recovered shear and IA estimates when using the FAO estimator. The vertical black lines show the mean recovered estimates using this method. The red curves show the distributed shear estimates when using the hybrid method, with the IA estimates identical for both methods. The vertical red lines show the mean recovered estimates using this method. The dashed green lines show the input signal. Here we see that both methods have successfully recovered shear and IA estimates with negligible bias. The dispersion in the FAO shear estimates is $\sim$40\% lower than those recovered using the CBB estimator and the dispersion in the IA estimates is $\sim$6\% lower for this set of input values. The dispersion in the hybrid shear estimates is $\sim$15\% lower than those recovered using the CBB estimator. The results are quantified in Table \ref{table:comp_ao}.}
\label{fig:disp_ao}
\end{minipage}
\end{figure*}
\begin{table*}
\begin{minipage}{6in}
\centering
\begin{tabular}{|c|c|c|c|c|c|c|c|c|}
\hline
Estimator $\left(\times10^{-2}\right)$ & $\sigma_{\hat{\gamma}_1}$ & $\sigma_{\hat{\gamma}_2}$ & $\left<\hat{\gamma}_1\right>$ & $\left<\hat{\gamma}_2\right>$ & $\sigma_{\hat{\gamma}_1^{\mathrm{IA}}}$ & $\sigma_{\hat{\gamma}_2^{\mathrm{IA}}}$ & $\left<\hat{\gamma}_1^{\mathrm{IA}}\right>$ & $\left<\hat{\gamma}_2^{\mathrm{IA}}\right>$ \\[0.5ex]
\hline
Angle-only & $1.02$ & $1.07$ & $-3.02\pm0.01$ & $4.02\pm0.01$ & $2.56$ & $2.57$ & $0.99\pm0.03$ & $-1.99\pm0.03$ \\
Hybrid & $1.42$ & $1.42$ & $-3.00\pm0.01$ & $4.01\pm0.01$ & $2.56$ & $2.57$ & $0.99\pm0.03$ & $-1.99\pm0.03$ \\ [1ex]
\hline
\end{tabular}
\caption{The mean and standard deviation of the shear and IA estimates
  recovered from $10^4$ simulations. Values are quoted for both the 
  angle-only estimator (equations (\ref{eq:est_aIA}) and (\ref{eq:corrected_F}), and equations (\ref{eq:est_atot}) - (\ref{eq:ao_shear})) and the hybrid
  estimator (equations (\ref{eq:est_aIA}) and (\ref{eq:corrected_F}), and equation (\ref{eq:hybrid})). The input shear and IA
  values are $\gamma_1=-0.03$, $\gamma_2=0.04$, $\gamma_1^{\mathrm{IA}}=0.01$ and $\gamma_2^{\mathrm{IA}}=-0.02$.}
\label{table:comp_ao}
\end{minipage}
\end{table*}
Assuming the same input values as used in Figure \ref{fig:disp_bb_cbb}, we recovered shear and IA estimates from $10^4$ realizations using the FAO estimator. The error on $\alpha^{\mathrm{int}}$ was assumed to be $10^{\circ}$ and the error on $\alpha^{\mathrm{obs}}$ was assumed to be zero to allow for a direct comparison of the performance of this estimator with the CBB estimator, where we assumed zero errors on the ellipticity measurements ($\bm{\epsilon}^{\mathrm{obs}}$). The results of this test are shown in Figure \ref{fig:disp_ao}. Note that the reduction in the dispersion of the shear and IA estimates is a result of the fact that we have assumed a perfect knowledge of the intrinsic ellipticity distribution. Errors on the prior knowledge of the intrinsic ellipticity distribution introduce multiplicative biases to the estimates of the shear and IA. This issue is addressed in \cite{whittaker14}, where constraints are placed on the size of the errors on the measurements of the ellipticities and the size of the sample used to estimate the intrinsic ellipticity distribution, such that this multiplicative bias is below a desired threshold value. Table \ref{table:comp_ao} shows the mean recovered shear and IA estimates and the standard deviation of the estimates.

A linear form of the estimator can be obtained by following the approach outlined in \cite{whittaker14}. Assuming that the $F_1\left(\left|\bm{\gamma}\right|\right)$ function can be approximated as
\begin{equation}\label{eq:lin_F}
F_1\left(\left|\bm{\gamma}\right|\right)\approx u\left|\bm{\gamma}\right|,
\end{equation}
for a general intrinsic ellipticity distribution we can find the coefficient, $u$, numerically. However, assuming a Rayleigh distribution for $\left|\bm{\epsilon}^{\mathrm{ran}}\right|$ and assuming that $\sigma_{\epsilon}$ is small enough for us to safely allow the limit in the integral, $\left|\bm{\epsilon}_{\mathrm{max}}^{\mathrm{int}}\right|$, to tend to infinity, it is possible to obtain the parameter, $u$, analytically. This is found to be
\begin{equation}\label{eq:u_rayleigh}
u=\left(\frac{\pi}{8\sigma_{\epsilon}^2}\right)^{\frac{1}{2}}.
\end{equation}
The linear approximation of the FAO estimator can then be written as
\begin{align}\label{eq:lin_ao_est}
\hat{\gamma}_1=&\frac{1}{uN}\sum_{i=1}^N\left[\frac{\cos\left(2\hat{\alpha}_i^{\mathrm{obs}}\right)}{\beta_2^{\mathrm{obs}}}-\frac{\cos\left(2\hat{\alpha}_i^{\mathrm{int}}\right)}{\beta_2^{\mathrm{int}}}\right], \nonumber \\
\hat{\gamma}_2=&\frac{1}{uN}\sum_{i=1}^N\left[\frac{\sin\left(2\hat{\alpha}_i^{\mathrm{obs}}\right)}{\beta_2^{\mathrm{obs}}}-\frac{\sin\left(2\hat{\alpha}_i^{\mathrm{int}}\right)}{\beta_2^{\mathrm{int}}}\right], \nonumber \\
\hat{\gamma}_1^{\mathrm{IA}}=&\frac{1}{uN}\sum_{i=1}^N\frac{\cos\left(2\hat{\alpha}_i^{\mathrm{int}}\right)}{\beta_2^{\mathrm{int}}}, \nonumber \\
\hat{\gamma}_2^{\mathrm{IA}}=&\frac{1}{uN}\sum_{i=1}^N\frac{\sin\left(2\hat{\alpha}_i^{\mathrm{int}}\right)}{\beta_2^{\mathrm{int}}}.
\end{align}
From here it is possible to recover an approximation for the dispersion in the estimator, given by
\begin{align}\label{eq:est_sig_ao}
\sigma_{\hat{\gamma}_1}=&\frac{1}{u\sqrt{N}}\left[\frac{1}{2\beta_2^{\mathrm{obs}^2}}+\frac{1}{2\beta_2^{\mathrm{int}^2}}-2\left<\cos\left(2\alpha^{\mathrm{obs}}\right)\cos\left(2\alpha^{\mathrm{int}}\right)\right>\right]^{\frac{1}{2}}, \nonumber \\
\sigma_{\hat{\gamma}_2}=&\frac{1}{u\sqrt{N}}\left[\frac{1}{2\beta_2^{\mathrm{obs}^2}}+\frac{1}{2\beta_2^{\mathrm{int}^2}}-2\left<\sin\left(2\alpha^{\mathrm{obs}}\right)\sin\left(2\alpha^{\mathrm{int}}\right)\right>\right]^{\frac{1}{2}}, \nonumber \\
\sigma_{\hat{\gamma}_1^{\mathrm{IA}}}=&\sigma_{\hat{\gamma}_2^{\mathrm{IA}}}=\frac{1}{u\sqrt{N}}\left[\frac{1}{2\beta_2^{\mathrm{int}^2}}\right]^{\frac{1}{2}}.
\end{align}
The dispersion in the shear estimates depends on the correlations between the cosines and sines of the true observed and intrinsic position angles. These, in turn, depend upon the intrinsic ellipticity distribution, the IA and the shear. In the absence of a shear signal, these correlation terms will, to first order in the IA signal, equal $1/2$. If we also neglect measurement errors, such that $\beta_2^{\mathrm{obs}}=\beta_2^{\mathrm{int}}=1$, then the dispersion in the shear estimates becomes zero. However, the presence of a non-zero shear signal reduces the correlation between the trigonometric functions and an error is introduced to the estimates. Hence, in the absence of measurement errors on the position angle measurements, the leading order term in the dispersion is dependent on the true shear. Measurement errors on the position angles also increase the dispersion in the estimates, as expected. Using the input values assumed in Figure \ref{fig:disp_ao} with equation (\ref{eq:est_sig_ao}), we recovered approximations for the errors on the shear and IA estimates by calculating the correlation terms numerically. The errors were found to be $\sigma_{\gamma_1}=1.06\times10^{-2}$, $\sigma_{\gamma_2}=1.13\times10^{-2}$ and $\sigma_{\gamma_1^{\mathrm{IA}}}=\sigma_{\gamma_2^{\mathrm{IA}}}=2.54\times10^{-2}$, which are in good agreement with the values quoted in Table \ref{table:comp_ao}.

Assuming that the shear signal is zero, we can also recover estimates of the dispersion in the shear estimates, which are $\sigma_{\gamma_1}=0.87\times10^{-2}$ and $\sigma_{\gamma_2}=0.85\times10^{-2}$. These values are approximately 20\% lower than the values quoted above where shear was included. Hence, we can conclude that the dispersion in the shear estimates depends strongly on the input shear signal, even if measurement errors on the position angles are included. This is an issue when trying to remove noise bias in power spectra estimates, and is discussed in more detail in Section \ref{sec:sims}. The dispersion in the IA estimates are independent of the true IA signal to first order.

\begin{figure*}
\begin{minipage}{6in}
\includegraphics{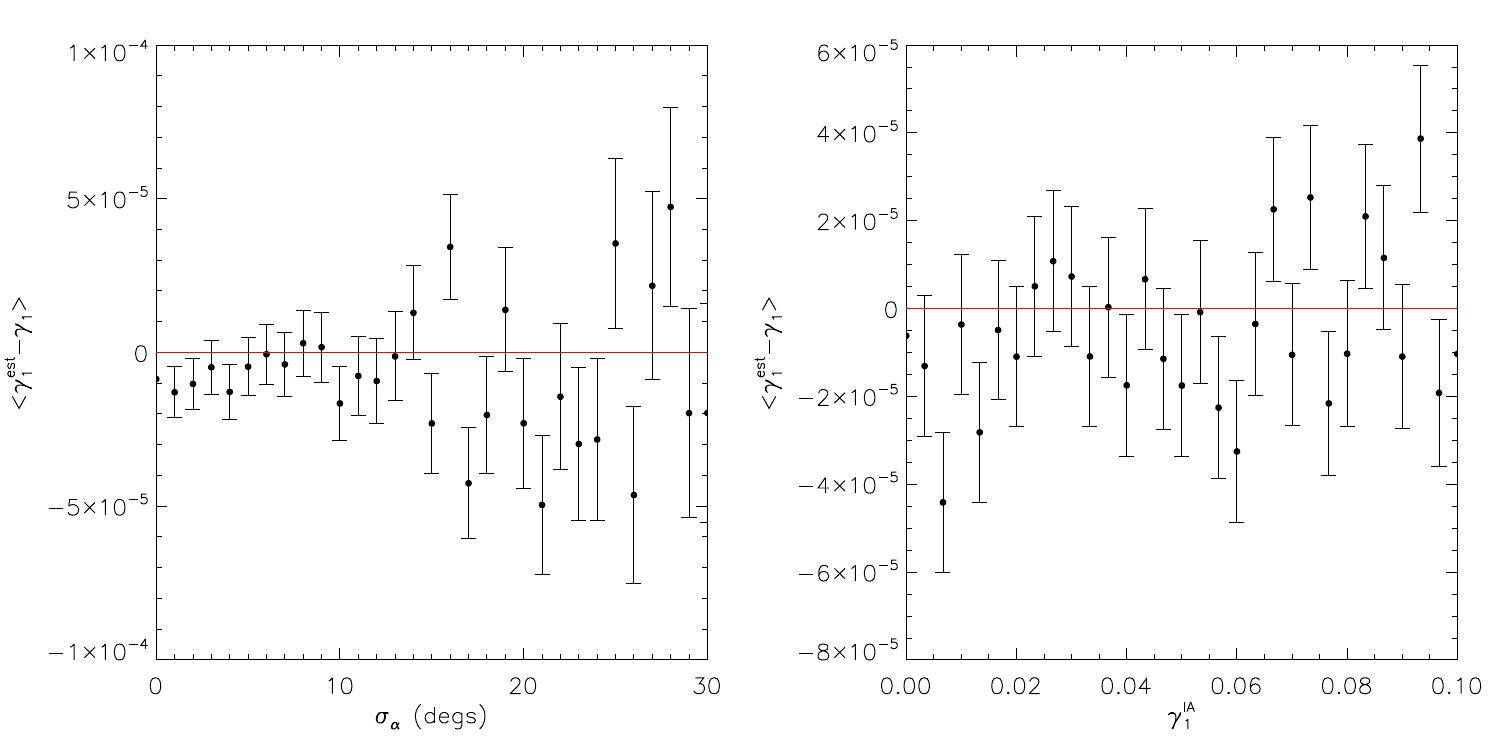}
\caption{Same as for Figure \ref{fig:bb_bias} but for the FAO shear estimator. From this plot see that any residual bias in the estimator can be considered negligible.}
\label{fig:ao_bias}
\end{minipage}
\end{figure*}

Figure \ref{fig:ao_bias} shows the residual bias in the FAO estimator. As with the CBB estimator, it is expected that there will be some residual bias from the nonlinear propagation of noise on the mean trigonometric functions into estimates of the shear. However, in all of the tests conducted this residual bias is found to be negligible.

\subsection{Hybrid Method}
\label{subsec:hyb}
In this subsection we introduce a hybrid method combining the standard method, which averages over galaxy ellipticity measurements, with the angle-only IA estimator.

Using a knowledge of the intrinsic ellipticity distribution, $f\left(\left|\bm{\epsilon}^{\mathrm{ran}}\right|\right)$, with measurements of the IPA only, we first recover an estimate of the IA signal via equations (\ref{eq:est_aIA}) and (\ref{eq:corrected_F}). We can then combine this estimate of the IA signal with an estimate of the vector $\bm{\gamma}+\bm{\gamma}^{\mathrm{IA}}$, provided by the mean of the observed ellipticities, to recover an estimate of the shear:
\begin{equation}\label{eq:hybrid}
\hat{\bm{\gamma}}=\left(\frac{1}{N}\sum_{i=1}^N\bm{\epsilon}_i^{\mathrm{obs}}\right)-\hat{\bm{\gamma}}^{\mathrm{IA}}.
\end{equation}

Using the same set of realizations as used to test the FAO method in Figure \ref{fig:disp_ao} (black curves), we recovered shear estimates from $10^4$ realizations using the hybrid shear estimator (equation (\ref{eq:hybrid})). The error on $\alpha^{\mathrm{int}}$ (which can be estimated using a measurement of the PPA) was assumed to be $10^{\circ}$ and the error on $\bm{\epsilon}^{\mathrm{obs}}$ was assumed to be zero. The results of this test are shown in Figure \ref{fig:disp_ao} as the red curves. It should be noted that, since the same realizations have been used to test the hybrid and FAO methods, the IA estimates are identical. As for the FAO estimator, discussed in Subsection \ref{subsec:ao}, the reduction in the dispersion of the shear estimates is a result of assuming a prior knowledge of the intrinsic ellipticity distribution when estimating the IA. Table \ref{table:comp_ao} shows the mean recovered shear and IA estimates and the standard deviation of the estimates.

Upon assuming a linear approximation of the $F_1\left(\left|\bm{\gamma}^{\mathrm{IA}}\right|\right)$ function using equation (\ref{eq:lin_F}), we can write a linear approximation of the hybrid shear estimator as
\begin{align}\label{eq:lin_hybrid}
\hat{\gamma}_1=&\frac{1}{N}\sum_{i=1}^N\left[\epsilon_{1,i}^{\mathrm{obs}}-\frac{\cos\left(2\hat{\alpha}_i^{\mathrm{int}}\right)}{u\beta_2^{\mathrm{int}}}\right], \nonumber \\
\hat{\gamma}_2=&\frac{1}{N}\sum_{i=1}^N\left[\epsilon_{2,i}^{\mathrm{obs}}-\frac{\sin\left(2\hat{\alpha}_i^{\mathrm{int}}\right)}{u\beta_2^{\mathrm{int}}}\right].
\end{align}
From here we can recover an approximation of the dispersion in the shear estimates:
\begin{align}\label{eq:est_sig_hybrid}
\sigma_{\hat{\gamma}_1}=&\frac{1}{\sqrt{N}}\left[\sigma_{\epsilon}^2+\frac{1}{2u^2\beta_2^{\mathrm{int}^2}}-\frac{2}{u}\left<\epsilon_1^{\mathrm{obs}}\cos\left(2\alpha^{\mathrm{int}}\right)\right>\right]^{\frac{1}{2}}, \nonumber \\
\sigma_{\hat{\gamma}_2}=&\frac{1}{\sqrt{N}}\left[\sigma_{\epsilon}^2+\frac{1}{2u^2\beta_2^{\mathrm{int}^2}}-\frac{2}{u}\left<\epsilon_2^{\mathrm{obs}}\sin\left(2\alpha^{\mathrm{int}}\right)\right>\right]^{\frac{1}{2}},
\end{align}
which depends upon the correlations between the true total ellipticities and the true intrinsic trigonometric functions. It can be shown that these correlation terms can be written as
\begin{align}\label{eq:cor_terms_hybrid}
\left<\epsilon_1^{\mathrm{obs}}\cos\left(2\alpha^{\mathrm{int}}\right)\right>=&\left<\epsilon_2^{\mathrm{obs}}\sin\left(2\alpha^{\mathrm{int}}\right)\right>\nonumber\\
\approx&u'+\mathcal{O}\left(\left|\bm{\gamma}\right|\left|\bm{\gamma}^{\mathrm{IA}}\right|\right)+\mathcal{O}\left(\left|\bm{\gamma}^{\mathrm{IA}}\right|^2\right),
\end{align}
where $u'$ is a zeroth order term, which is independent of the input shear and IA signals, but is dependent on the form of the intrinsic ellipticity distribution, $f\left(\left|\bm{\epsilon}^{\mathrm{ran}}\right|\right)$. Hence, to first order in the shear and IA, the correlation terms are constant, and therefore, we can approximate the dispersion in the shear estimates to be
\begin{equation}\label{eq:est_sig_hybrid_final}
\sigma_{\hat{\gamma}_1}\approx\sigma_{\hat{\gamma}_2}\approx\frac{1}{\sqrt{N}}\left[\sigma_{\epsilon}^2+\frac{1}{2u^2\beta_2^{\mathrm{int}^2}}-\frac{2u'}{u}\right]^{\frac{1}{2}}.
\end{equation}
For a Rayleigh distribution it is possible to recover the coefficient $u'$ analytically, if we adopt the same assumptions used to derive equation (\ref{eq:u_rayleigh}). This is found to be
\begin{equation}\label{eq:udash_rayleigh}
u'=\left(\frac{\pi\sigma_{\epsilon}^2}{8}\right)^{\frac{1}{2}}.
\end{equation}
Therefore, we can conclude that, in the absence of shear and IA signals, and assuming zero measurement errors on the estimates of $\alpha^{\mathrm{int}}$, there is a dispersion in the shear estimates which arises from random shape noise. With these assumptions, we found in the previous subsection that the dispersion in the FAO shear estimator was zero. For the case of the FAO estimator, a knowledge of the intrinsic ellipticity distribution is assumed for the random component of both the observed and intrinsic ellipticities. This allows us to recover estimates of the vectors $\left(\bm{\gamma}+\bm{\gamma}^{\mathrm{IA}}\right)$ and $\bm{\gamma}$ using only measurements of $\hat{\alpha}^{\mathrm{obs}}$ and $\hat{\alpha}^{\mathrm{int}}$. Using only measurements of the position angles eliminates the contribution of random shape noise in the FAO shear estimates. However, the hybrid estimator requires measurements of $\bm{\epsilon}^{\mathrm{obs}}$, which contributes random shape noise to the estimates of the shear. This noise is, to first order, independent of both the shear and IA signals. 

Using equation (\ref{eq:est_sig_hybrid_final}) with the input values used in Figure \ref{fig:disp_ao}, we recovered approximations for the dispersion in the shear estimates using the hybrid method. These were found to be $\sigma_{\hat{\gamma}_1}=\sigma_{\hat{\gamma}_2}=1.40\times10^{-2}$, which are in good agreement with the values quoted in Table \ref{table:comp_ao}. 

\begin{figure*}
\begin{minipage}{6in}
\includegraphics{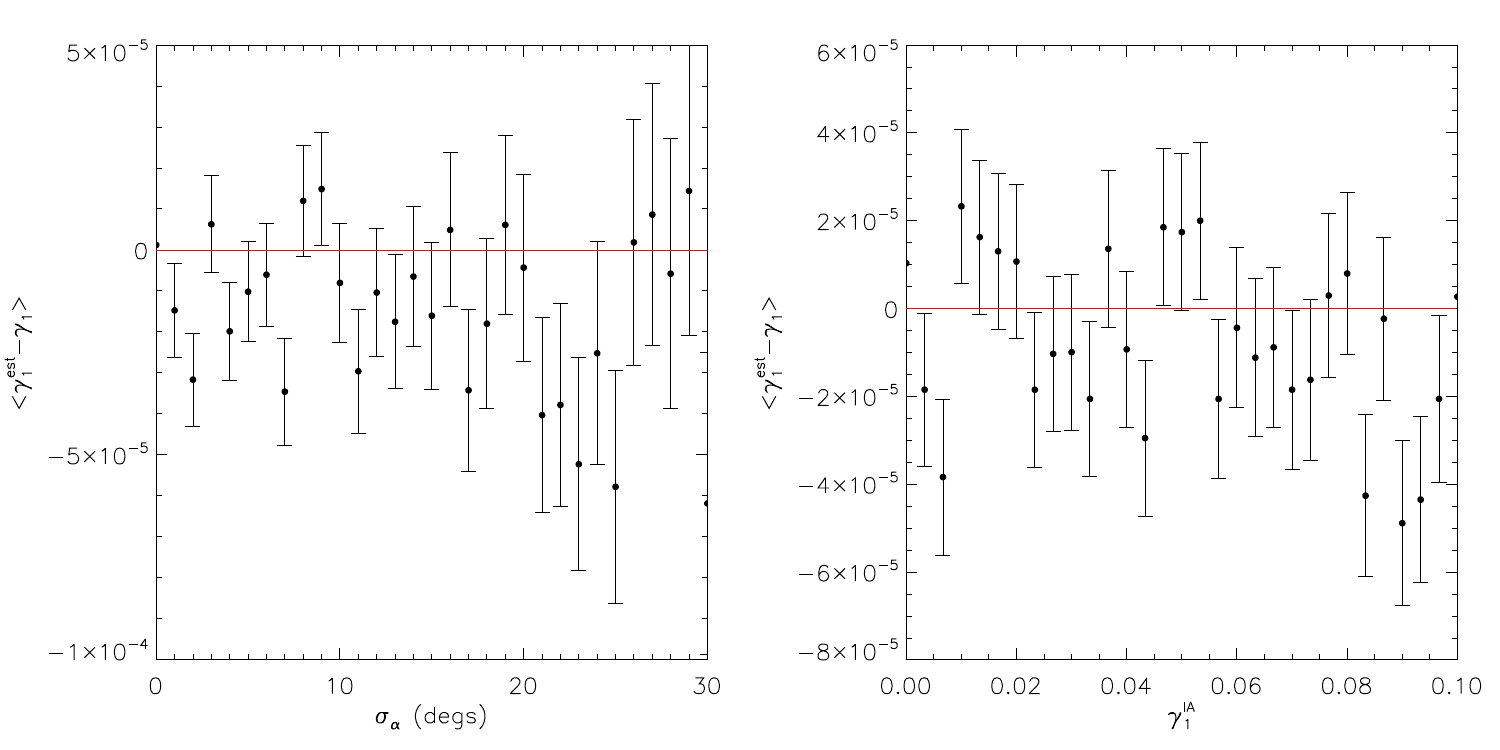}
\caption{Same as for Figure \ref{fig:bb_bias} but for the hybrid shear estimator. Here we see a small residual bias due to the finite number of source galaxies. However, this bias is much smaller than the bias in the original BB estimator (Figure \ref{fig:bb_bias}).}
\label{fig:hybrid_bias}
\end{minipage}
\end{figure*}

Figure \ref{fig:hybrid_bias} shows the residual bias in the hybrid estimator. Here we see a residual bias which is a result of the nonlinear propagation of noise on the mean trigonometric functions into estimates of the IA signal and hence into the shear estimates. However, we find that this bias is much smaller than the dispersion in the estimates in all of the tests we have conducted, and is negligible when we consider the power spectra reconstructions in Section \ref{sec:sims}.

\section{Tests on Simulations}
\label{sec:sims}
In this section we test the three estimators, described in the previous sections, by reconstructing the lensing and IA auto and cross-power spectra following the approach described in BB11. All of the simulated fields are assumed to be pure Gaussian fields and, as our aim is to demonstrate the power of the estimators to separate the shear and IA signals given an unbiased estimate of the intrinsic position angle, we ignore the effects of observational systematics.

In all simulations we assume a $\Lambda$CDM background cosmology, with the matter density parameter $\Omega_{\mathrm{m}}=0.262$, the amplitude of density fluctuations $\sigma_8=0.798$, the Hubble constant $H_0=71.4\,\mathrm{km}\,\mathrm{s}^{-1}\,\mathrm{Mpc}^{-1}$, the baryon density parameter $\Omega_{\mathrm{b}}=0.0443$ and the scalar spectral index $n_{\mathrm{s}}=0.962$.

We simulate the weak lensing and IA fields in three different redshift bins and include all of the possible cross-correlations between the fields in the different bins. The selected bins are $0.00<z_1<1.40$, $1.40<z_2<2.60$ and $z_3>2.60$, with the bin limits selected such that each bin contains approximately the same number density of sources. We simulate the IA signal using the modified non-linear alignment model introduced by \cite{bridle07} and we use a normalization for the IA power spectrum which is five times the observed SuperCOSMOS level in order to make it easier to see the effects we are dealing with, and we assume a correlation coefficient of $\rho_c=-0.2$. A detailed discussion of the simulated fields is given in BB11. Here we focus on the performance of the estimators.

In order to demonstrate the methods discussed, we assume that all of the galaxies have sufficient information to measure the IPA. We assume a measurement error on the $\alpha^{\mathrm{int}}$ (IPA) estimates with r.m.s. $10^{\circ}$ and a negligible error on the ellipticity and $\alpha^{\mathrm{obs}}$ measurements in order to make a fair comparison between the various methods.

To estimate the power spectra from the simulated observations, we pixelize the sky into $3.4\times3.4\text{ arcmin}^2$ cells and assume a background galaxy number density of $4\text{ arcmin}^{-2}$ for each redshift bin; this number density is chosen to avoid the issue discussed at the end of Subsection \ref{subsec:req_num}, though we note that this may be achievable in future deep surveys with the SKA. We then reconstruct shear and IA maps using each of the estimators discussed in the previous sections. 

We estimate the recovered power spectra from the reconstructed shear and IA maps using the standard pseudo-$C_l$ approach (\citealt{brown11a, hivon02, brown05}). 

In the presence of noise in the shear estimates, we can write the general expectation value of the estimated pseudo-$C_l$ power spectra, $\tilde{C}_l^{XY}$, as
\begin{equation}\label{eq:pseudo_cl}
\left<\tilde{C}_l^{XY}\right>=C_l^{X_sY_s}+C_l^{X_nY_n}+C_l^{X_sY_n}+C_l^{X_nY_s},
\end{equation}
where the postscripts $X$ and $Y$ denote the fields being correlated and where the subscripts $s$ and $n$ respectively denote the signal and noise in that field. One can correct for biases due to noise and correlations between the signal and noise, such that an unbiased estimate of the power spectra can be recovered using a suite of Monte-Carlo simulations:
\begin{equation}\label{eq:pseudo_est_spec}
\hat{C}_l^{XY}=\tilde{C}_l^{XY}-\left<C_l^{X_nY_n}\right>_{\mathrm{mc}}-\left<C_l^{X_sY_n}\right>_{\mathrm{mc}}-\left<C_l^{X_nY_s}\right>_{\mathrm{mc}},
\end{equation}
where the angle brackets indicate the mean over the suite of Monte-Carlo simulations. This is the form of the power spectra estimator used for the remainder of this paper. In the presence of model dependent noise and correlations between the signal and noise, unbiased estimates of the power spectra are only achievable if the Monte-Carlo simulations include the input power spectra. In a real analysis, this will obviously not be possible. In order to address this issue, we adopt an iterative approach to estimating the spectra. To begin with, we construct a suite of 200 Monte-Carlo simulations under the assumption that the input shear and IA signals are zero. This provides us with an initial estimate of the power spectra using equation (\ref{eq:pseudo_est_spec}). As we shall see this is sufficient when using the CBB and hybrid methods to recover the shear and IA estimates. However, it is insufficient when using the FAO estimator. It does, however, provide us with initial estimates of the power spectra. These initial estimates can then be used to construct a suite of improved Monte-Carlo simulations, which can be used to update our estimates of the power spectra.

\begin{figure*}
\begin{minipage}{6in}
\includegraphics{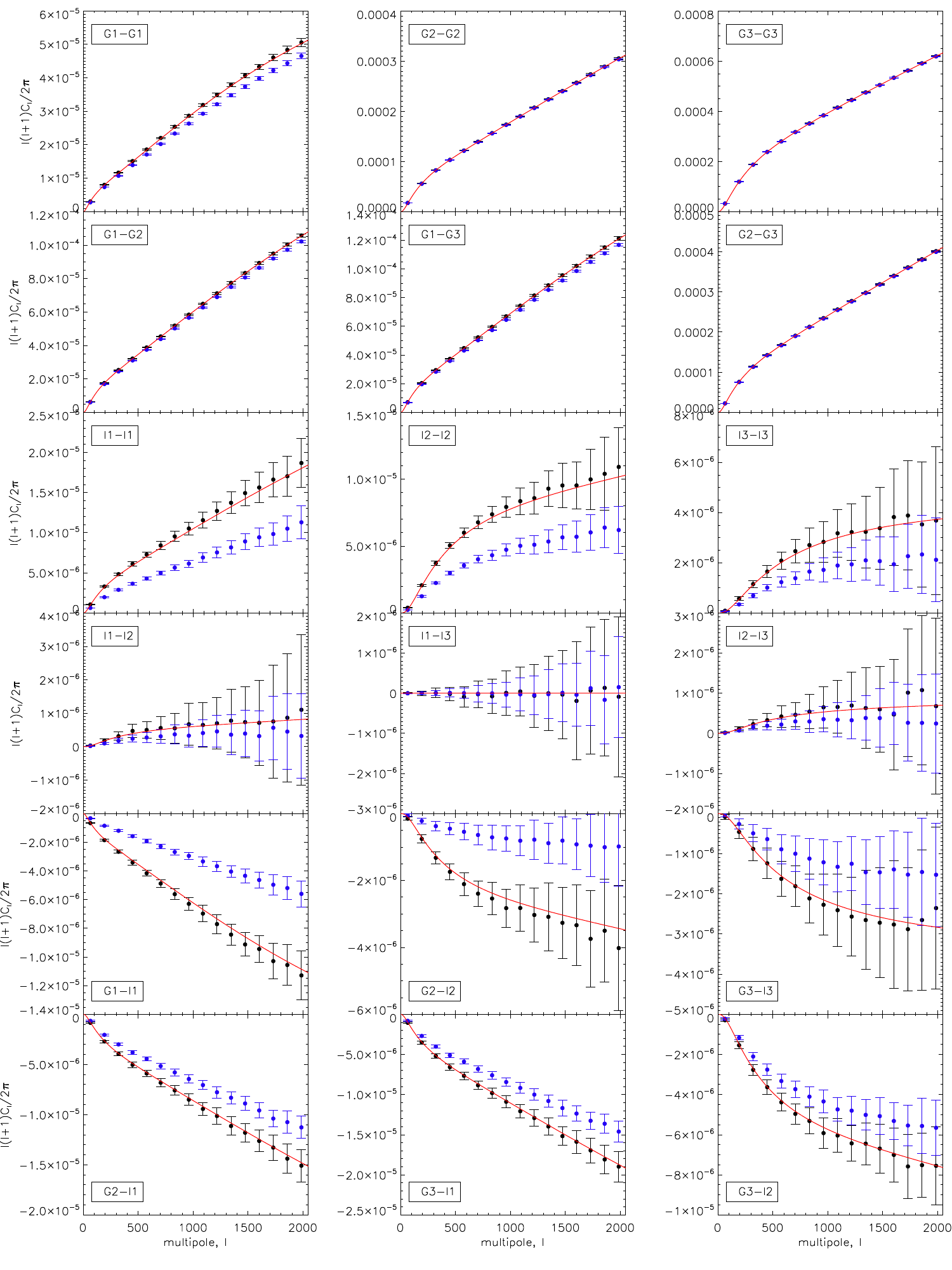}
\caption{Reconstructions of the lensing and IA auto and cross-power spectra. In each panel the red curve shows the model power spectra. The black points show the reconstructed power spectra using the CBB estimator to estimate the shear and IA signals. The blue points show the reconstructions using the original BB estimator, as a comparison. From these reconstructions we clearly see that the residual bias has been reduced when using the CBB estimator.}
\label{fig:ps_cbb_bb}
\end{minipage}
\end{figure*}

\begin{figure*}
\begin{minipage}{6in}
\includegraphics{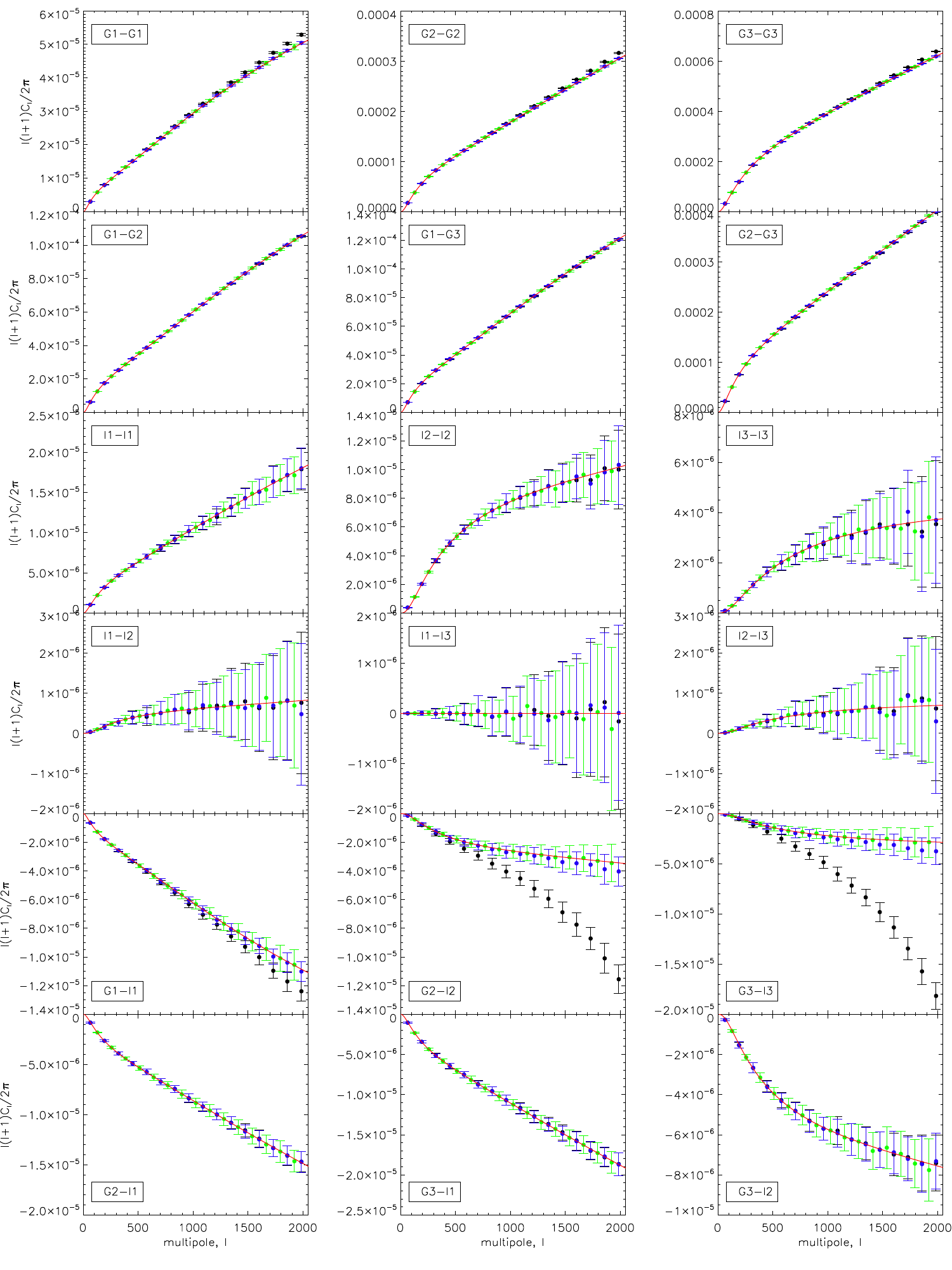}
\caption{Reconstructions of the lensing and IA auto and cross-power spectra. In each panel the red curve shows the model power spectra. The black points show the reconstructed power spectra using the FAO estimator to estimate the shear and IA signals and with the noise maps created under the assumption that the input shear and IA signals are zero. The blue points show the reconstructed power spectra recovered upon using the iterative procedure described in the main text. From this we see the success of the iterative procedure. The green points show the reconstructed power spectra using the hybrid estimator to estimate the shear and IA signals, with no iterative procedure required.}
\label{fig:ps_ao_stao}
\end{minipage}
\end{figure*}

Figure \ref{fig:ps_cbb_bb} shows the reconstructed shear and IA auto and cross-power spectra for each of the three overlapping redshift bins, recovered using the CBB estimator (black points) and using a suite of 200 Monte-Carlo simulations under the assumption that the input shear and IA signal are zero. The blue points show the reconstructed power spectra using the original BB estimator to estimate the shear and IA. The red curves show the input power spectra. From this we clearly see the success of the correction.

Figure \ref{fig:ps_ao_stao} shows the reconstructed power spectra when using the FAO estimator (black points). The linear form of the FAO estimator, given in equation (\ref{eq:lin_ao_est}), has been used to reduce computation time. From this we see that there is a residual bias in the shear power spectra which propagates into estimates of the shear-IA cross-power spectra. This bias is due to a dependence of the errors on the shear estimates on the input shear signal, as described at the end of Subsection \ref{subsec:ao}. This bias is not successfully corrected for when using noise-only Monte-Carlo simulations. However, if we use these estimated power spectra as the input power spectra for a further set of Monte-Carlo simulations, we can construct noise and noise-signal power spectra which include an estimate of the shear and IA signal. These updated noise and noise-signal power spectra can then be used to recover an improved estimate of the input power spectra. This step can be iterated until subsequent estimates of the power spectra are deemed consistent. The blue points in Figure \ref{fig:ps_cbb_bb} are the result of this procedure, using just one iteration. We see that the iterative step has, indeed, improved our estimates of the power spectra.

The green points in Figure \ref{fig:ps_ao_stao} show the reconstructed power spectra using the hybrid estimator. The linear form of the hybrid estimator, given in equation (\ref{eq:lin_hybrid}), has been used to reduce computation time. This reconstruction did not require the use of the iterative procedure. It was shown, in the discussion which follows from equation (\ref{eq:est_sig_hybrid_final}), that the dispersion in the shear estimates is independent of the input shear and IA signal to first order. Hence, the zero-signal noise power spectra successfully removes the noise bias from the power spectra without the need of the iterative procedure described above.

\begin{figure*}
\begin{minipage}{6in}
\includegraphics{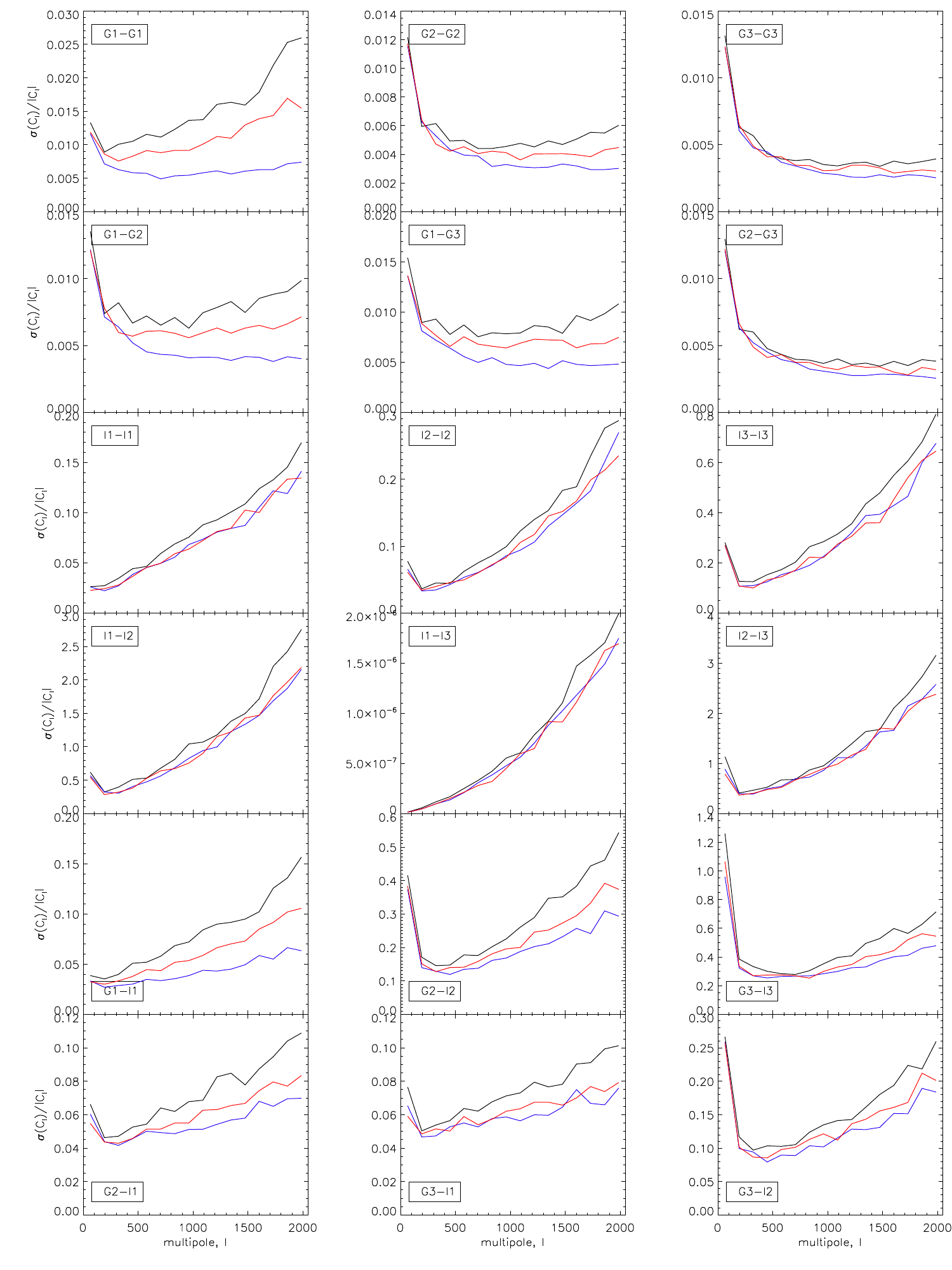}
\caption{The fractional error in the power spectra reconstructions shown in Figures \ref{fig:ps_cbb_bb} and \ref{fig:ps_ao_stao}. The curves show the CBB estimator (black), the FAO estimator (blue) and the hybrid estimator (red). Note, the input I1-I3 cross-power spectrum is zero for all multipoles and therefore we show the error as opposed to the fractional error for that panel.}
\label{fig:frac_err}
\end{minipage}
\end{figure*}

\begin{table*}
\begin{minipage}{6in}
\centering
\begin{tabular}{|c|c|c|c|c|}
\hline
Spectrum & Original BB & Corrected BB & FAO & Hybrid \\ [0.5ex]
\hline
G1-G1 & $(-7.48\pm0.02)\times10^{-2}$  & $(7.45\pm0.29)\times10^{-3}$ & $(1.45\pm0.13)\times10^{-3}$ & $(0.17\pm0.21)\times10^{-3}$ \\
G2-G2 & $(-3.99\pm0.10)\times10^{-3}$  & $(9.60\pm1.14)\times10^{-4}$ & $(3.90\pm0.96)\times10^{-4}$ & $(1.77\pm1.05)\times10^{-4}$ \\
G3-G3 & $(-1.97\pm0.10)\times10^{-3}$  & $(4.02\pm1.01)\times10^{-4}$ & $(-6.27\pm0.93)\times10^{-4}$ & $(0.69\pm0.98)\times10^{-4}$ \\
G1-G2 & $(-3.07\pm0.01)\times10^{-2}$  & $(2.44\pm0.16)\times10^{-3}$ & $(-0.38\pm0.11)\times10^{-3}$ & $(0.26\pm0.13)\times10^{-3}$ \\
G1-G3 & $(-3.48\pm0.01)\times10^{-2}$  & $(2.68\pm0.17)\times10^{-3}$ & $(-0.83\pm0.12)\times10^{-3}$ & $(-0.05\pm0.15)\times10^{-3}$ \\
G2-G3 & $(-4.73\pm0.10)\times10^{-3}$  & $(6.42\pm1.03)\times10^{-4}$ & $(-11.72\pm0.95)\times10^{-4}$ & $(0.79\pm0.99)\times10^{-4}$ \\
I1-I1 & $(-3.85\pm0.01)\times10^{-1}$  & $(3.15\pm0.17)\times10^{-2}$ & $(0.17\pm0.14)\times10^{-2}$ & $(-0.27\pm0.14)\times10^{-2}$ \\
I2-I2 & $(-3.85\pm0.02)\times10^{-1}$  & $(3.80\pm0.38)\times10^{-2}$ & $(-0.12\pm0.29)\times10^{-2}$ & $(-0.39\pm0.30)\times10^{-2}$ \\
I3-I3 & $(-3.83\pm0.09)\times10^{-1}$  & $(-0.66\pm1.57)\times10^{-2}$ & $(2.79\pm1.33)\times10^{-2}$ & $(2.31\pm1.42)\times10^{-2}$ \\
I1-I2 & $(-4.04\pm0.19)\times10^{-1}$  & $(1.15\pm0.32)\times10^{-1}$ & $(-0.56\pm0.27)\times10^{-1}$ & $(0.33\pm0.27)\times10^{-1}$ \\
I2-I3 & $(-4.60\pm0.30)\times10^{-1}$  & $(1.85\pm0.49)\times10^{-1}$ & $(-0.09\pm0.42)\times10^{-1}$ & $(-0.23\pm0.42)\times10^{-1}$ \\ [1ex]
\hline
\end{tabular}
\caption{The mean fractional bias in the power spectra reconstructions across all multipoles.}
\label{table:ps_mean_frac_bias}
\end{minipage}
\end{table*}

In Figure \ref{fig:frac_err} we show the fractional errors on the reconstructed power spectra. From this we see, as expected, that the errors are largest for the CBB estimator, and smallest for the FAO estimator. However, we emphasize that we have assumed a perfect knowledge of $f\left(\left|\bm{\epsilon}^{\mathrm{ran}}\right|\right)$ when using the FAO and hybrid estimators. This would obviously not be the case in a real analysis where uncertainties on our knowledge of $f\left(\left|\bm{\epsilon}^{\mathrm{ran}}\right|\right)$ would lead to an increase in the errors of the FAO and hybrid approaches.

Table \ref{table:ps_mean_frac_bias} shows the mean fractional bias in the shear auto and cross-power spectra and the IA auto and cross-power spectra reconstructions.
From this we see that there is approximately an order of magnitude reduction in the fractional bias of the CBB estimator, as compared with the original BB estimator. The reduction in bias is generally greater when using the FAO and hybrid methods. However, these methods require an accurate knowledge of $f\left(\left|\bm{\epsilon}^{\mathrm{ran}}\right|\right)$ and, for the case of the FAO estimator, an iterative method to remove noise bias.

\section{Conclusions}
\label{sec:conclusion}
When we include a correction term into the formalism of the estimator introduced by BB11, we have demonstrated that the residual bias in the estimator, which emerges in the presence of measurement errors on the intrinsic position angle estimates and a non-zero IA signal, can be reliably reduced to negligible levels as compared with the original BB estimator, provided that a sufficient number of resolved background galaxies have reliable polarization information. When including the correction term, chance alignments of the measured IPA may result in substantial outliers in the distribution of the shear estimates if the number of background galaxies is small. However, we have introduced a method which may be used to place constraints on the number of background galaxies required to recover reliable shear estimates when using this estimator. This restriction may require large cells, such that the number of source galaxies within each cell is greater than or equal to the minimum number of galaxies required, and hence small scale information may not be attainable.

Building upon the angle-only estimator introduced by \cite{whittaker14}, we have constructed an angle-only IA estimator which uses IPA measurements and requires a knowledge of the intrinsic ellipticity distribution. From here we can formulate two distinct shear estimators. The first is the FAO shear estimator, which requires measurements of the observed position angles. The second is the hybrid method, which combines the angle-only IA estimator with the standard shear estimator and requires measurements of the observed ellipticities. We have demonstrated that both of these methods may be used to recover shear estimates which exhibit negligible residual biasing as compared with the original BB estimator. The FAO method, however, requires the implementation of an iterative procedure to mitigate the effects of a signal dependent noise bias in the shear and shear-IA power spectra. We further emphasize that the results presented in this paper are based on the assumption that the distribution, $f\left(\left|\bm{\epsilon}^{\mathrm{ran}}\right|\right)$, is known exactly. An incorrect knowledge of this distribution propagates as a multiplicative bias into the shear and IA estimates. However, it is expected that this distribution may be accurately measured using deep calibration observations in future surveys. Constraints on the accuracies required, and the number of galaxies required to achieve these accuracies, are discussed in \cite{whittaker14}. 

Present radio surveys, such as SuperCLASS, currently under observation using the JVLA and e-MERLIN\footnote{http://www.e-merlin.ac.uk/legacy/projects/superclass.html} arrays, will hopefully provide information about the fraction of galaxies with reliable polarization information and the expected error on the intrinsic orientation estimates, $\sigma_{\alpha^{\mathrm{int}}}$, provided by measurements of the PPA. This information is essential if we are to gain an understanding of the cosmological scales which may be probed using these techniques. In addition to this, we hope to improve our understanding of the impact of Faraday rotation on measurements of the PPA. It is expected that this effect may be corrected for using information from multiple frequencies to extract the rotation measures of the source galaxies. 

We aim to apply these techniques to future radio surveys, such as with the SKA, where the number density of galaxies will be higher than for current radio surveys, enabling us to probe smaller cosmological scales. The high redshifts achieved by the SKA will also enable radio weak lensing to probe regions of the Universe which are inaccessible to other weak lensing surveys \citep{brown15}. This high redshift information will provide powerful constraints on the evolution of large-scale structure in the Universe. Another exciting prospect for future radio weak lensing is the cross-correlation of radio and optical weak lensing surveys, such as the correlation of shear estimates from EUCLID with those from the SKA. This method has the advantage that the systematics in the two telescopes are expected to be completely uncorrelated, allowing the effects of systematics to be removed from shear analyses, while avoiding the residual effects of an incorrect calibration.

\renewcommand{\theequation}{A-\arabic{equation}}
\renewcommand{\thefigure}{A-\arabic{figure}}
\setcounter{equation}{0}  
\setcounter{figure}{0}
\section*{Appendix A: Correcting the Brown \& Battye Estimator}
\label{cor_bb}
In this section we discuss the correction to the BB estimator. 

We begin by redefining the matrix $\mathbf{A}$ as
\begin{equation}\label{eq:redmat_A}
\mathbf{A}=\frac{2\beta_4^{\mathrm{int}}}{N}\sum_{i=1}^Nw_i\hat{\boldsymbol{n}}_i\hat{\boldsymbol{n}}_i^T,
\end{equation}
and the vector $\bm{b}$ as
\begin{equation}\label{eq:redvec_b}
\boldsymbol{b}=\frac{2\beta_4^{\mathrm{int}}}{N}\sum_{i=1}^Nw_i\left(\boldsymbol{\epsilon}^{\mathrm{obs}}_i\cdot\hat{\boldsymbol{n}}_i\right)\hat{\boldsymbol{n}}_i,
\end{equation}
where the definition of the vector $\hat{\bm{n}}_i$ is given in equation (\ref{eq:pseudodef}) and where $w_i$ is a normalized arbitrary weight assigned to each galaxy. In the limit $N\rightarrow\infty$, the matrix $\mathbf{A}$ can be written as
\begin{equation}\label{eq:matrix_A_subcorr}
\mathbf{A}=\beta_4^{\mathrm{int}}\left(
\begin{array}{cc}
1 - \left<\cos\left(4\hat{\alpha}^{\mathrm{int}}\right)\right> & -\left<\sin\left(4\hat{\alpha}^{\mathrm{int}}\right)\right> \\
-\left<\sin\left(4\hat{\alpha}^{\mathrm{int}}\right)\right> & 1 + \left<\cos\left(4\hat{\alpha}^{\mathrm{int}}\right)\right>
\end{array} \right),
\end{equation}
and the vector $\bm{b}$ can be written as
\begin{equation}\label{eq:vector_b_subcorr}
\bm{b}=\beta_4^{\mathrm{int}}\left(
\begin{array}{c}
\left<\epsilon_1^{\mathrm{obs}}\left[1-\cos\left(4\hat{\alpha}^{\mathrm{int}}\right)\right]\right>-\left<\epsilon_2^{\mathrm{obs}}\sin\left(4\hat{\alpha}^{\mathrm{int}}\right)\right> \\
\left<\epsilon_2^{\mathrm{obs}}\left[1+\cos\left(4\hat{\alpha}^{\mathrm{int}}\right)\right]\right>-\left<\epsilon_1^{\mathrm{obs}}\sin\left(4\hat{\alpha}^{\mathrm{int}}\right)\right>
\end{array} \right).
\end{equation}

In the presence of noise on the estimates of $\alpha^{\mathrm{int}}$, the trigonometric functions above will be biased. This bias can be corrected for by dividing the functions by the correction term $\beta_4^{\mathrm{int}}$, as defined in equation (\ref{eq:gen_beta}), such that the corrected matrix $\mathbf{A}$ can be written as
\begin{equation}\label{eq:matrix_A_corr}
\mathbf{A}=\left(
\begin{array}{cc}
\beta_4^{\mathrm{int}} - \left<\cos\left(4\hat{\alpha}^{\mathrm{int}}\right)\right> & -\left<\sin\left(4\hat{\alpha}^{\mathrm{int}}\right)\right> \\
-\left<\sin\left(4\hat{\alpha}^{\mathrm{int}}\right)\right> & \beta_4^{\mathrm{int}} + \left<\cos\left(4\hat{\alpha}^{\mathrm{int}}\right)\right>
\end{array} \right),
\end{equation}
and the corrected vector $\bm{b}$ becomes
\begin{equation}\label{eq:vector_b_corr}
\bm{b}=\left(
\begin{array}{c}
\left<\epsilon_1^{\mathrm{obs}}\left[\beta_4^{\mathrm{int}}-\cos\left(4\hat{\alpha}^{\mathrm{int}}\right)\right]\right>-\left<\epsilon_2^{\mathrm{obs}}\sin\left(4\hat{\alpha}^{\mathrm{int}}\right)\right> \\
\left<\epsilon_2^{\mathrm{obs}}\left[\beta_4^{\mathrm{int}}+\cos\left(4\hat{\alpha}^{\mathrm{int}}\right)\right]\right>-\left<\epsilon_1^{\mathrm{obs}}\sin\left(4\hat{\alpha}^{\mathrm{int}}\right)\right>
\end{array} \right).
\end{equation}

The CBB estimator can then be written more concisely by defining the matrix $\mathbf{M}_i$, where 
\begin{equation}
\mathbf{M}_i=\left(
\begin{array}{cc}
\beta_4^{\mathrm{int}} - \cos\left(4\hat{\alpha}_{i}^{\mathrm{int}}\right) & -\sin\left(4\hat{\alpha}_{i}^{\mathrm{int}}\right) \\
-\sin\left(4\hat{\alpha}_{i}^{\mathrm{int}}\right) & \beta_4^{\mathrm{int}} + \cos\left(4\hat{\alpha}_{i}^{\mathrm{int}}\right)
\end{array} \right),
\end{equation}
such that the final form of the estimator can be written as described by equations (\ref{eq:correctBB}) - (\ref{eq:h_matrix}).

\section*{Acknowledgments}
LW and MLB are grateful to the ERC for support through the award of an
ERC STarting Independent Researcher Grant (EC FP7 grant number 280127). MLB also thanks the STFC for the award of Advanced and
Halliday fellowships (grant number ST/I005129/1). 
\bibliographystyle{mn2e}
\bibliography{ms}

\label{lastpage}

\end{document}